\documentclass[leqno,12pt]{article}
\usepackage{cite}
\usepackage[a4paper,hscale=0.82,vscale=0.75]{geometry}
\usepackage{graphicx}
\usepackage{amssymb}

\usepackage{amsmath}
\usepackage{amsthm}
\usepackage{amscd}
\usepackage[all,cmtip]{xy} 
\numberwithin{equation}{section}
\usepackage{empheq}
\usepackage[british]{babel}
\usepackage[small,nohug]{diagrams} 
\usepackage{enumitem}
\setlist[enumerate]{label=\arabic*$^\circ$,leftmargin=*,before=\setlength{\rightmargin}{\leftmargin}}

\title{Degree of arbitrariness directly from moving frames}
\author{Ziyang Hu\footnote{\texttt{z.hu@damtp.cam.ac.uk}}\\
D.A.M.T.P.\\
 University of Cambridge}

\newcommand{\pd}{\partial}

\newtheoremstyle{shape0}
  {9pt}
  {9pt}
  {}
  {}
  {\bfseries}
  {.}
  {.5em}
  {}

\newtheorem*{ckt}{Cartan--K\"ahler Theorem}
\newtheorem*{cit}{Cartan's Involutivity Test}
\newtheorem*{cit2}{Cartan's Involutivity Test, Version 2}
\newtheorem*{cpt}{Theorem on the complete prolongation of Pfaffian systems}

\newtheoremstyle{shape1}
  {9pt}
  {9pt}
  {}
  {\parindent}
  {\it}
  {.}
  { }
  {}

\newtheoremstyle{shape2}
  {9pt}
  {9pt}
  {}
  {}
  {\itshape}
  {.}
  {.5em}
  {}

\theoremstyle{shape1}
\newtheorem{rmk}{Remark}
\newtheorem*{hnt}{Historical Note}

\theoremstyle{shape0}

\theoremstyle{shape2}
\theoremstyle{definition}

\DeclareMathOperator{\iprod}{\lrcorner}

\newcommand{\str}{^{\star}}

\begin{document}
\maketitle
\abstract{For physical theories, the degree of arbitrariness of a system is of great importance, and is often closely linked to the concept of degree of freedom, and for most systems this number is far from obvious. In this paper we present an easy to apply algorithm for calculating the degree of arbitrariness for large classes of systems formulated in the language of moving frames, by mere manipulation of the indices of the differential invariants. We then give several examples illustrating our procedure, including a derivation of the degree of arbitrariness of solutions of the Einstein equation in arbitrary dimensions, which is vastly simpler than previous calculations, the degree of arbitrariness of gauge theories of arbitrary groups under Yang--Mills type equations, for which until now only specific cases have been calculated by rather messy means, and the degree of arbitrariness of relativistic rigid flow with or without additional constraints, which has not been derived before. Finally we give the proof of our algorithm.}

\newpage

\tableofcontents
\newpage

\section{Introduction}

\subsection{The question of the degree of arbitrariness of a system}

The degree of arbitrariness of physical systems is defined as the number of arbitrary functions that one must specify in order to completely determine the state of the system. Often it is also closely linked to the concept of the degree of freedom of a system, which is heuristically described as the number of ways the constituents of the system can move independently: for example, the degree of freedom in the kinetic theory of gases, or of elementary particles. The concept of degree of arbitrariness is general and not restricted to microscopic systems nor systems constructed with classical particles: we also talk about the degree of arbitrariness of a solid body, of a fluid flow, of the electromagnetic field, of a gauge field, or even of the gravitational field. As physical theories are written in terms of equations, the importance of the degree of arbitrariness, interpreted as the number of arbitrary functions entering the solution of the equations of the system, cannot be overemphasised.

However, despite the great importance of this quantity, there is so far no general strategy for calculating it. In the classical theory of gases or of rigid bodies, calculation is done mostly ``by intuition'', and in slightly more complicated circumstances, \emph{ad hoc} arguments and calculations are used to suggest a degree of arbitrariness (feasible since one is almost always restricted to dimension $3+1$), which is then justified \emph{a posteriori} by checking that the suggested number is physically sensible. In the theory of fields, the degree of arbitrariness, when it is mentioned at all, is almost always calculated ``by definition'': the geometrical space of the theory itself is constructed using a number of free functions. These strategies work, either because the system is fairly simple, or the construction of the system is straightforward. But even in the latter case there are often many pitfalls: for example, the metric of spacetime has $6$ degrees of arbitrariness, but a Riemannian geometry is constructed using a bilinear form, which involves $10$ free functions. The usual argument goes that we must take into the ``diffeomorphism invariance'' of spacetime into account, which subtracts $4$ from the counting. Of course, such arguments are a bit hand-waving, and whenever a geometrical setting is defined, instead by postulating something like the metric, by specifying a connection, such arguments do not apply.
 Related to Riemannian geometry, we also talk about the degree of arbitrariness of Einstein's gravity, which is, roughly speaking two times the degree of freedom of the graviton: remember that when calculating the degree of freedom of something, we are really restricting to a Cauchy surface for the equations of motion, and a function together with its time derivative, which are really two independent functions on the surface, counts as a single degree of freedom.

In more complicated systems, especially in systems that are defined by using one or more \emph{constraints}, the above-mentioned strategies fail miserably. Consequently, when faced with such a theory, instead of having a clear picture of what the \emph{physical} variables are, we deal with a large number of dependent variables linked together by interrelated partial differential equations. One can still do calculations with such systems. However, due to the ignorance of the independent quantities in such systems, sooner or later calculations will have the tendency of degenerating into the random walk of some drunken sailor.

In this paper, we propose an algorithm for calculating the degree of arbitrariness of a system by geometrical considerations. 
Our algorithm works by first formulating the physical problems using the method of moving frames.
As physical theories are often \emph{generic} in nature: functions are not explicitly given, and are required to have their \emph{invariance properties}: two systems are considered to be equivalent if a allowed transformation transforms one to the other, the use of the method of moving frames immediately grants us an incredible advantage to our algorithm: as the method of moving frames uses no coordinates whatsoever, the degree of arbitrariness we obtain is the true, physical degree of arbitrariness: there is no business of things like ``diffeomorphism invariance of the coordinates'', or ``gauge invariance'', etc., as we shall see later \footnote{It should be noted that the method of moving frames is \emph{not} the same as the tetrad method widely used in physics: the tetrad method depends on coordinates: it just derives a new set of coordinates from old ones. The method of moving frames, on the other hand, use differential forms, instead of coordinate functions, as the fundamental vehicle for doing calculations.}. It is hoped that the proposed method will promote the method of moving frames, which has hitherto seen various physical applications in formulating the kinematics of physical systems, to be used also in studying the dynamics of physical systems, since it provides a way of studying equations of motions and constraints without specifying any coordinates.

As physical theories are ultimately formulated using differential equations, any calculation of degrees of arbitrariness amounts to the study of solutions of differential equations. Our algorithm is no different, and at its roots it relies on the Cauchy--Kowalewski existence theorem for systems of partial differential equations. However, that everything we do is reducible to the Cauchy--Kowalewski theorem does not mean that once we know this theorem we know everything about calculating degrees of arbitrariness for physical theories. There are two problems: first, as we have mentioned, physical theories are often given in a generic form (which may even have the dimension of spacetime as a variable), and without the explicit functional forms of the variables, the Cauchy--Kowalewski theorem is not directly applicable; second, even if a theory is explicitly constructed, a direct application of the Cauchy--Kowalewski theorem would involve too much work. Our algorithm, on the other hand, makes use of an already non-trivial consequence of the Cauchy--Kowalewski theorem, the Cartan--K\"ahler existence theorem, to devise a simple procedure that can be applied without understanding any of the underlying mechanisms. If nothing else, simplicity should be the merit of our algorithm: one is required only to know how to do index manipulations and apply simple reasonings in combinatorics in order to apply our algorithm.

\subsection{The historical origin of our algorithm}

Our algorithm is based on the method of moving frames, the Cartan--K\"ahler theory of exterior differential systems, and to a lesser extent, the method of equivalence, all of which are due to the French mathematician \'Elie Cartan. Indeed, Cartan himself became briefly interested in the problem of calculating the degrees of arbitrariness of physical equations, and specifically those of the equations of general relativity and Einstein's (failed) unified field theory. With the equations explicitly given, Cartan of course knows how to calculate the degree of arbitrariness (see the historical note on page \pageref{ba}), and he communicated the methods he used to Einstein in a series of letters. At first suspicious, Einstein finally understood Cartan's methods, and he wrote in his reply to Cartan \cite{cartan-einstein}:
\begin{quote}
  Ich habe Ihr Manuskript gelesen und zwar \emph{mit Begeisterung}. Nun ist mir alles klar. Sie sollten diese Theorie ausflihrlich publizieren, denn ich glaube, dass sie fundamentale Bedeutung hat.

(I have read your manuscript --- \emph{enthusiastically}. Now everything is clear to me. You should publish this theory in detail; I believe it is of fundamental importance.)
\end{quote}
At Einstein's suggestion, Cartan did publish his methods with physicists in mind as his audience \cite{cartan-phy}. However, perhaps for fear that the physicists will not be too comfortable with his method of moving frames, which is still being actively developed by Cartan in private at the time, he chose to formulate the method so that it applies to first order partial differential equations, and due to the arbitrariness of the coordinates (and for modern applications, gauge ``degrees of arbitrariness'') involved, and the fact that his formulation requires  knowledge of all the relations among the quantities in the differential equations, which are easy to be missed out \emph{unless} one is using the method of moving frames, his method is difficult to apply, and even Einstein made various mistakes in applying the method \cite{cartan-einstein}. 
After Einstein gave up his unified theory of fields, the Cartan--Einstein correspondences stopped, and Cartan never again attempted to communicate his methods to the physical community. It is very sad his published method did not spread very far among physicists, both due to the difficulty of the method and Einstein's later isolation from the mainstream physics community.

Our algorithm can be considered a modern and much easier to apply reformulation of Cartan's original methods. The key improvements made to the original method is the consistent use of moving frames and the theory of differential invariants, both of which are also developed by Cartan.  Our method also applies in cases where the mere form, instead of the explicit dimensions, etc., of a physical problem, are given.

In Cartan's days it is still quite rare, and usually deemed unnecessary, to think in higher dimensions or arbitrary dimensions. In his last book \cite{cartan-eds}, Cartan gave many examples of applications of his theory of exterior differential systems to obtain degrees of arbitrariness of many geometrical problems by direct application of the Cartan--K\"ahler theorem, some of them quite relevant for physics. But since most of these problems takes place on familiar dimensions (dimension $2$ or $3$), there is not too much motivation for generalisations or trying to find common features in the calculations. It is perhaps also for this reason that Cartan did not even attempt to formulate an easier method for calculating degrees of arbitrariness, and there is no doubt that he would have done much better than the present author did if he lives to see this day.

\subsection{Structure of this paper}

After a brief discussion of the necessary settings, we will immediately state our algorithm in section \ref{sec:alg} without proof.  Then in section \ref{sec:ex}, we give several example applications of our algorithms. Several examples chosen are such that the degree of arbitrariness is well-known, as a check of the correctness of our algorithm. Historical notes are included with these problems. The examples also include some new results, including the degree of arbitrariness of a general Yang--Mills type theory, and the degree of arbitrariness of relativistic rigid flow under various conditions. For more involved applications of the algorithm to more complex problems, see the papers by the author \cite{hu1,hu2}. Finally, in section \ref{sec:pr} we give the proof of our algorithm.

The rather unusual and not very logical ordering we chose for our presentation is due to the consideration that the algorithm itself is simple enough to understand and apply to physical problems without too much additional mathematical baggages, whereas the proof involves the Cartan--K\"ahler theory on exterior differential systems, which few physicists know about. 
 The mathematical constructions and theorems on which our algorithm is based on are reviewed in two appendices in this paper: we choose not to point simply to a book of reference since some of these books are rather dense and have to be read in a linear fashion. In addition, the modern approach to the theory is often highly formal, whereas we prefer a more down-to-earth approach suitable for calculations. As a consequence, our presentation of the Cartan--K\"ahler theory and of the equivalence method is closer to the original presentation of Cartan \cite{cartan-eds,cartan-grp,cartan-pf} than to what is usually done today.

\section{The algorithm}
\label{sec:alg}

We will first state our algorithm. Our description aims to be precise, and as a consequence the statements in the following may seem long-winded and complicated. This is only because we try to avoid ``heuristics'' in our statements, and as our examples will show later, once a problem has been formulated in the language of moving frames, the difficulty of the applications of our algorithm is only that of rather simple problems in the combinatorics of the indices.

\subsection{The moving frame}

As we have mentioned, first of all we must place the system we want to study into the language of moving frames. Assume this has been done: thus let $\omega_{\mu}$, $\mu=1,2,\dots,N$ be a co-frame for a suitable manifold $P$ constructed from the geometrical problem. We can then proceed to write down the \emph{structural equations}
\begin{equation}
  \label{eq:1}
  d\omega_{\mu}=I_{\mu\nu\lambda}\omega_{\nu}\wedge\omega_{\lambda},
\end{equation}
where we have employed the Einstein summation convention. The quantities $I_{\mu\nu\lambda}$, which are scalar functions on $P$, are called the \emph{fundamental invariants} of the system. In addition, we may have a set of functions $J_{\alpha}$ subject to certain constraints. We further assume that we can find a subset of the 1-forms, denoted by $\omega_{i}$, $i=1,2,\dots,n$, such that all $\omega_{\mu}$, $I_{\mu\nu\lambda}$ and $F_{\alpha}$ are labelled using indices $i,j,k\dots$ instead of $\mu,\nu,\rho\dots$, or those labelling the additional functions.

From now on we will treat the additional functions $J_{\alpha}$ (which may or may not be given together with some explicit functional dependence among themselves and the $I_{\mu\nu\lambda}$) as additional fundamental differential invariants, and unless explicitly mentioned, ``fundamental invariants'' means both $I_{\mu\nu\lambda}$ and $J_{\alpha}$.

\begin{rmk}
  It is perfectly possible to construct $\omega_{\mu}$ from
  coordinates. But once these forms have been constructed, we should
  treat them as fundamental and forget the coordinates. All
  geometrical properties of the system will be derived through the
  invariants $I_{\mu\nu\lambda}$. This is the whole point of the
  method of moving frame. 
\hfill\P\end{rmk}

\begin{rmk}
  We can take the indices $i,j,k\dots$ to run over all values of
  $\mu,\nu,\lambda\dots$, and then the assumption of the existence of
  this subset of indices will be valid for all systems. The utility of
  our assumption, on the other hand, will be clear in a moment.
\hfill\P\end{rmk}

\begin{rmk}
In cases where we have additional functions $J_{\alpha}$, these are considered further constraints: two systems are considered equivalent if and only if we can set $\omega_{\mu}=\omega_{\mu}'$ and $J_{\alpha}=J'_{\alpha}$.
The incorporation of the possibility of additional function into our algorithm gives us great flexibility in application of our theory. For example, if for example, we have a system with one-forms $\omega_{\mu}$ spanning the cotangent space but are not independent, we can choose a subset of these one forms as the co-frame and the rest can be written in terms of the co-frame with the help of additional functions, and thus this system can be treated by our algorithm. On the other hand, if a system has certain one-forms $\omega_{\mu}$ which do not span the cotangent space, we can choose arbitrarily some other one-forms in order to form a coframe, and then prolong the problem to obtain a coframe on a suitable principal bundle, \cite{JOlver:1995p9015} has more details on this. Thus in this case our algorithm is also applicable.
\hfill\P\end{rmk}

\subsection{The derived invariants}

We can differentiate the fundamental invariants:
\[
dI_{\mu\nu\lambda}=I_{\mu\nu\lambda;\delta}\omega_{\delta},
\]
where the functions $I_{\mu\nu\lambda;\delta}$ are called \emph{first order derived invariants}. As $\omega_{\mu}$ form a co-frame, they are uniquely determined in terms of the co-frame. Derivations can be carried out further:
\[
dI_{\mu\nu\lambda;\delta}=I_{\mu\nu\lambda;\delta\gamma}\omega_{\gamma},
\]
where the functions $I_{\mu\nu\lambda;\delta\gamma}$ are called \emph{second order derived invariants}. Derived invariants of all order are defined recursively.

By our assumption, the one-forms $\omega_{\mu}$ other than $\omega_{i}$ are labelled using indices $i,j,k\dots$, and we denote such forms by $\omega_{I}, \omega_{J}, \dots$ where $I,J\dots$ are multi-indices in $i,j,k\dots$. Thus, for any fundamental or derived invariant $I_{I;J}$, we have
\[
dI_{I;J}=I_{I;Jk}\omega_{k}+I_{I;J;K}\omega_{K},
\]
which is nothing more than the above formulae rewritten using different indices. We assume that the derived invariants $I_{I;J;K}$ can be expressed linearly in terms of $I_{I;Jk}$ and lower order invariants.

\begin{rmk}
  As before, if we take $i,j,k\dots$ to run over all possible indices,
  then the above assumption is trivially verified since there is then
  no $I_{I;J;K}$. In almost all applications, however, the
  $\omega_{i}$ will be the \emph{horizontal} one-forms in a principal bundle and the above
  assumption means that only the horizontal derivatives can be
  independent, i.e., we have a \emph{connection} at work here. The
  derived quantities $I_{I;Jk}$ are then the \emph{covariant
    derivatives} of $I_{I;J}$ in the bundle if the bundle is reductive \cite{WSharpe:1997p5521}.
\hfill\P\end{rmk}

\subsection{Algebraic relations}

Next we need to take into account algebraic relations of the invariants. 
When we write down the structural equations, the fundamental invariants appearing in the structural equations are subject to certain relations (symmetries). For example, for the most general structural equation \eqref{eq:1}, the symmetry is $I_{\mu\nu\lambda}=-I_{\mu\lambda\nu}$. When we stipulate the additional functions, they may also be subject to certain relations. We call such algebraic relations the \emph{fundamental algebraic relations}.

We can derive the structural equations: 
\[
0=d^{2}\omega_{\mu}=dI_{\mu\nu\lambda}\wedge\omega_{\nu}\wedge\omega_{\lambda}+I_{\mu\nu\lambda} d\omega_{\nu}\wedge d\omega_{\lambda}-I_{\mu\nu\lambda}\omega_{\nu}\wedge d\omega_{\lambda},
\]
and, after using the structural equations themselves and the resolution of $d I_{\mu\nu\lambda}$ in terms of derived invariants, we obtain
\[
0=F_{\mu\nu\lambda}\omega_{\mu}\wedge\omega_{\nu}\wedge\omega_{\lambda},
\]
where $F_{\mu\nu\lambda}$ are functions in the differential invariants. We call the algebraic relations
\[
F_{[\mu\nu\lambda]}=0
\]
the \emph{Bianchi algebraic relations}.

If $I_{I;J}$ is a differential invariant, deriving it we get
\[
dI_{I;J}=I_{I;Jk}\omega_{k}+I_{I;J;K}\omega_{K},
\]
and deriving again we get
\[
0=I_{I;Jkl}\omega_{k}\wedge\omega_{l}+C_{IJkl}\omega_{k}\wedge\omega_{l}+A_{IJAk}\omega_{A}\wedge\omega_{k}+B_{IJAB}\omega_{A}\wedge\omega_{B},
\]
where the functions $C_{IJkl}$ contains only invariants of lower order. We make the further assumption that $A_{IJAk}=0$ and $B_{IJAB}=0$ identically. 
\begin{rmk}
  Again, if $i,j,k\dots$ runs over all values the assumption is
  trivially verified. If, as we have mentioned, the choice of indices
  $i,j,k\dots$ comes from the existence of a connection, this
  assumption can also be verified easily.
\hfill\P\end{rmk}
The relations
\[
I_{I;Jkl}=I_{IJlk}+C_{IJkl}
\]
are called the \emph{generic algebraic relations}.

Obviously, if an invariant $I_{I}$ is actually one of the additional functions in the system, there will not be any Bianchi algebraic relation for it.

If $R=0$ is an algebraic relation, we can derive it to obtain
\[
dR=R_{i}\omega_{i}+R_{A}\omega_{A}=0,
\]
and we have the new relations
\[
R_{i}=0, \qquad R_{A}=0
\]
which are called the \emph{derived algebraic relations}.

\begin{rmk}
  As before, the relation $R_{A}=0$ is usually either vacuous or an
  identity.
\hfill\P\end{rmk}

The defining relations, the Bianchi relations, the generic relations and their derived relations are all the algebraic relations of the invariants.

\subsection{The involutive seeds and the degree of arbitrariness}
\label{sec:invseeds}
First we need to define the concept of \emph{covering} of invariants. In the following, when we say that a function $f$ is expressible using the functions $g_{\alpha}$, $\alpha\in I$ where $I$ is some indexing set, we mean that we can find an \emph{explicit} function $\tilde f$ such that
\[
f=\tilde f(g_{\alpha_{1}},g_{\alpha_{2}},\dots,g_{\alpha_{k}}),\qquad \alpha_{1},\alpha_{2},\dots,\alpha_{k}\in I.
\]
In particular, if $f$ is a constant, then it is expressible using the empty set.

Let $S$ be a set of invariants (the covering set) and $T$ be another set of invariants (the set to be covered). From $S$, we form the set $S'$ by
\begin{enumerate}
\item adding all elements expressible using the elements of $S$ to $S'$
\item adding all elements $I_{I;J}$ such that $I_{I;Jk}$ for all $k$ can be expressed in terms of elements in $S$,
\end{enumerate}
and we repeat this procedure, using $S'$ in place of $S$, to obtain $S''$, etc. This procedure may eventually stabilise. In practice, due to the way we use the set $S^{(k)}$, we only need to add terms that are independent: see below.

Suppose at a certain stage we have the set $S^{(k)}$. We remove all elements of $T$ that can be expressed in terms of elements in $S^{(k)}$ to form a set $\bar T^{(k)}$. 
If every element $J_{J;K}$ of $\bar T^{(k)}$ are such that $J_{J;KL}$ for all multi-indices $L$ of a certain length $l$ are expressible in terms of the empty set, i.e., are constants, then $S$ covers $T$. In particular, $S$ covers $T$ if $\bar T^{(k)}$ is empty.


This definition may seem a little long-winded and peculiar, especially the part dealing with non-empty $\bar T^{(k)}$. As a result of this definition, if a set of invariants $I_{I;J}$ has $I_{I;JL}=$ constant for all multi-indices $L$ of a certain length, then it is covered by the empty set.

A system of \emph{involutive seeds} is a set of differential invariants that satisfies the following conditions:
\begin{enumerate}[label=\textbf{I\arabic*}.]
\item (Covering.) All invariants occurring in the structural equations $\omega_{\mu}=\cdots$ and the derived structural equations $d\omega_{\mu}=\cdots$, as well as in any algebraic relations to be enforced, are covered;
\item (Independence.) All invariants in the set are algebraically independent, namely there is no non-trivial relations among them, possibly also involving invariants that are covered by the set of involutive seeds, that is derivable from the algebraic relations that we mentioned. Hence, no proper subset of this set is a covering of the above-mentioned invariants.
\item (Derivation index.) Every invariant in the set contains at least one derivation index.  (This condition can be non-vacuous only for invariants coming from additional functions.)
\end{enumerate}

Now we will give the indices $i,j,k\dots,n$ an ordering labelled by the first few natural numbers, i.e., a function $s$ such that $s(i)\in \mathbb{N}$, which amounts to a permutation, or relabelling, of the numerical indices if the indices themselves run from $1$ to some number without gaps. An ordering will be called an \emph{involutive ordering} if the following two conditions are satisfied:
\begin{enumerate}[label=\textbf{O\arabic*}.]
\item (Maximality of lower indices.) If two invariants $I_{Ii}$ and $J_{Jj}$ occur in a relation for some value of $i$ and $j$ and $i>j$, then $I_{Ii}$ must \emph{not} be an involutive seed.
\item (Counting condition.) For every involutive seed $I_{ijk\dots l}$, for all indices $m < l$ using this ordering, the invariants $I_{ijk\dots m}$ are also involutive seeds. (Note: they must occur explicitly, not merely expressible as linear combinations of others.)
\end{enumerate}
The names given to the conditions appearing in parentheses above will become clear when we prove our algorithm.

\begin{rmk}
That an ordering we write down is an
  involutive ordering with respect to a system of involutive seeds is the
  \emph{test} of our algorithm, and we will obtain information about the degree of arbitrariness only if this test passes. In
  the next section we will show, by examples, of how to proceed in
  order to have a good chance of arriving at an involutive ordering.
\hfill\P\end{rmk}

We are now nearly done: armed with an involutive ordering for a system of involutive seeds, in many cases we can read off the degree of arbitrariness of the system directly. Let us describe one further test:
  \begin{enumerate}[label=\textbf{R}.]
  \item (Rank condition.) We can find a number of algebraically independent
    invariants $I_{\alpha}$, $\alpha=1,2,\dots$, which are \emph{not}
    additional functions introduced and which when derived give
    \[
    dI_{\alpha}=C_{\alpha\mu}\omega_{\mu},
    \]
    and the rank of the matrix $C_{\alpha\mu}$ is equal
    to the number of one-forms $\omega_{\mu}$.
  \end{enumerate}
Assume that this test is satisfied. Let $S'$ be the subset of $S$ such that all elements are such that their last index is maximal with respect to the involutive ordering among elements of the sets. Let this index be $d$ and let $k$ be the number of elements of $S'$. Then we say that the system has degree of arbitrariness 
 $k$, occurring at dimension $s(d)$. If the set of involutive seeds is the empty set (this is possible, since we have seen that a set can be covered by the empty set), then the system does not have any degree of arbitrariness. It may or may not be inconsistent.

\begin{rmk}
  Another way of phrasing the same result is that the system depends
  on
  $k$ independent, arbitrary functions of $s(d)$
  variables. See section \ref{sec:pr}.
\hfill\P\end{rmk}

Granted, there are systems for which the test \textbf{R} fails. For such systems, we can still apply our algorithms, but the number we obtained is only meaningful if the system does not involve additional functions that we put in by hand, and still the number is only guaranteed to be an \emph{upper bound} of the degree of arbitrariness of the system.

\begin{rmk}
  As we will see, if the test \textbf{R} fails, then the question we are posing are not formulated on the best space possible: we can reduce the problem into a lower dimensional one, for which the test \textbf{R} holds and we can obtain the degree of arbitrariness, not merely an upper bound. If \textbf{R} fails and we have additional functions defined on the space, we may be able to use the additional functions to specialise the moving frames so as to reduce the dimension. Note that such a reduction of dimension need not be a reduction of the dimension of the \emph{base manifold}: for example, in the principal bundle over a manifold Riemannian, we may be able to reduce the bundle from an $SO(n)$ bundle to a suitable $G$ bundle where $G$ is a subgroup of $SO(n)$, and in extreme cases $G$ may even be a discrete group.

As any well-formulated problem will satisfy the test \textbf{R}, most often the verification of this condition is easy.\hfill\P
\end{rmk}




\subsection{Adding ``equations of motion''}

After we have obtained the degree of arbitrariness coming from the geometry of a system, we have effectively solved parts of the so-called \emph{Cauchy problem} for the system: the involutive seeds that have their last index maximal are those that we can specify arbitrarily, which, together with suitable boundary conditions, will determine the system uniquely.

However, these functions that we can arbitrarily specify are not the only possible choice, and often they are not the functions that we really want to specify: most of the time they are derivatives of too high an order. Usually what we really want to do is to specify some functions of the low order invariants. The question is then: can we specify \emph{these} functions arbitrarily?

In a guise, this is precisely the problem of adding equations of motion to a geometrical (kinematical) system. The equations of motion are usually differential equations, but as higher order invariants are nothing more than derivatives of the fundamental equations, these equations of motion are, in our framework, just additional algebraic relations. Hence, by taking these equations as additional defining relations for the differential invariants, our method is still applicable: the new degree of arbitrariness, after the introduction of equations of motion, signify the \emph{Cauchy data} for the equations of motion.

Note that after the introduction of additional relations the old system of involutive seeds and ordering may not apply at all: for example, a completely different ordering may be needed. The new system may also fail to admit any system of involutive seeds and ordering: this is in particular the case if the equations of motion introduced is potentially contradictory. These will all be illustrated by examples later.

\begin{rmk}
  Real, physical equations of motion have additional nice properties: they always reduce the dimension at which the degree of arbitrariness occurs by exactly one. See the examples later.\hfill\P
\end{rmk}

\subsection{The directions of evolution}

There is one additional bit of information that we can obtain from our manipulation of indices. Assume that for a system, our algorithm applies with the condition \textbf{R} satisfied. Then for a system of involutive seeds together with its involutive ordering, let $S$ be the set of last indices of the involutive seeds, and let $T$ be the set of all possible indices. Then the set of indices $T-S$ gives the \emph{directions of evolution} of the problem. The interpretation is as follows: if we consider the Cauchy problem of the system, then the system is specified completely by specifying $d$ functions on a $k$ dimensional submanifold of the manifold in which the system is defined, where $d$ is the degree of arbitrariness and $k$ is the dimension at which the degree of arbitrariness occurs. But in general, this submanifold cannot be chosen at will: it must be transverse to the system of vectors having indices taken from the set $T-S$ for a certain choice of involutive seeds and ordering (this choice is in general not unique). As a consequence of our assumptions that the one-forms splits into two sets $\omega_{i}$ and $\omega_{\alpha}$ and all independent invariants take the indices of $\omega_{i}$, we see that all directions corresponding to $\omega_{\alpha}$ are automatically directions of evolution. If the space for which the system is defined is a principal bundle and $\omega_{i}$ are the horizontal forms whereas $\omega_{\alpha}$ are the vertical forms, this just affirms the fact that the data in the bundle is completely determined once we specify the data on a section of the bundle.

There is a further constraint on the choice of the submanifold: it must not contain the so-called characteristic directions. For this constraint, see the proof of our algorithm.

\begin{rmk}
  For real, physical equations of motion, the single additional direction of evolution when compared with the system without the equations added must be the time direction.\hfill\P
\end{rmk}

\subsection{A little adjustment in case of failure}

There are cases for which our algorithm is not applicable, since we will not be able to find any system of involutive seeds and involutive orderings. Most often, the problem is of the following kind: we have two invariants $I_{I;Jk}$ and $I'_{I';J'k;}$, and for $k=a$, $I_{I;Ja}$ is independent and $I'_{I';J'a}$ is not, whereas for $k=b$, $I'_{I';J'b}$ is independent whereas $I_{I;Jb}$ is not. Thus we can take neither $a>b$ nor $a<b$ for our involutive ordering.

In most such case, the solution is easy: we just take the forms $\omega_{a}$ and $\omega_{b}$ and replace them with $\omega'_{a}=\alpha\omega_{a}+\beta\omega_{b}$ and $\omega_{b}'=\gamma\omega_{a}+\delta\omega_{b}$, with
\[
\det
\begin{pmatrix}
  \alpha&\beta\\
  \gamma&\delta
\end{pmatrix}
\neq 0,
\]
and it is easy to show that, for this new co-frame, which is otherwise completely equivalent to our old one, we can either take $I_{I;Ja}$ and $I'_{I';J'a}$ to be independent, or $I_{I;Jb}$ and $I'_{I';J'b}$.
Of course, for complicated systems involving a large number of offending one-forms, this might get quite complicated.

\section{Examples}
\label{sec:ex}

If the description of the algorithm above seems too abstract, the examples below will show how easy it is to apply it and how little the work required is.

\subsection{Riemannian geometry and Einstein gravity}

A Riemannian geometry is defined locally by specifying a non-degenerate, positive definite symmetric bilinear form. ({The assumption of positive definiteness is not required for what follows and is added only to simplify the exposition. However, there will be differences when we discuss the equations of motion: see the remarks later.}) In dimension $n$, such a bilinear form has $n(n+1)/2$ independent components. By a not very precise argument, since two Riemannian metrics $g$ and $g'$ on $M$ and $M'$ are considered equivalent if there is a function $f:M\rightarrow M'$ such that $f^{*}(g')=g$, the real degree of arbitrariness is the number of independent components less the number of coordinates, namely $n(n-1)/2$.

In the moving frame formulation, a Riemannian geometry on $M$ is locally defined by a principal bundle $\pi:P\rightarrow M$, which is locally the same as $SO(n)\times M$. A suitable coframe on $P$ is given by the system of horizontal forms $\omega_{i}$ and the vertical forms $\omega_{ij}$, with $\omega_{ij}$ forming a representation of the Lie algebra $\mathfrak{so}(n)$. The structural equations are
\begin{equation}
  \label{eq:rstreq}
\left\{
  \begin{aligned}
    d\omega_{i}&=-\omega_{ij}\wedge\omega_{j},\\
    d\omega_{ij}&=-\omega_{ik}\wedge\omega_{kj}+\tfrac{1}{2}R_{ijkl}\omega_{k}\wedge\omega_{l},
  \end{aligned}
\right.
\end{equation}
which can also be written in the matrix form:
\[
d\omega=-\omega\wedge\omega+\Omega,
\]
where
\[
\omega=
\begin{pmatrix}
  0&0\\
  \omega_{i}&\omega_{ij}
\end{pmatrix},\qquad
\Omega=
\begin{pmatrix}
  0&0\\
  0&R_{ijkl}\omega_{k}\wedge\omega_{l}
\end{pmatrix},
\]
making the Lie algebra structure explicit. The fundamental invariants are $R_{ijkl}$, with defining symmetries
\begin{equation}
  \label{eq:def}
  R_{ijkl}=-R_{jikl},\qquad R_{ijkl}=-R_{ijlk},
\end{equation}
the first coming from the fact that $\omega_{ij}$ is a representation of $\mathfrak{so}(n)$, and the second due to how $R_{ijkl}$ appears in the structural relation. Obviously  we want to take $\omega_{i}$ as the basic forms, as everything else is labelled using their indices. The definition of the first derived invariants
\begin{equation}
  \label{eq:def1stinv}
  dR_{ijkl}=R_{ijkl;m}\omega_{m}-R_{mjkl}\omega_{im}-R_{imkl}\omega_{jm}-R_{ijml}\omega_{km}-R_{ijkm}\omega_{lm}
\end{equation}
shows that our first assumption about the basic forms is satisfied, and $R_{ijkl;m}$ are really the covariant derivatives of $R_{ijkl}$. The definition of higher order invariants are straightforward.

For the Bianchi relations, $d^{2}\omega_{i}=0$ gives
\begin{equation}
  \label{eq:bianchi1}
  R_{i[jkl]}=0,
\end{equation}
whereas $d^{2}\omega_{ij}=0$ gives
\begin{equation}
  \label{eq:bianchi2}
  R_{ij[kl;m]}=0,  
\end{equation}
i.e., they are just the usual Bianchi identities. The generic relations and the derived relations are obvious.

For this system, the invariants $R_{ijkl;m}$ cover all the invariants occurring in the structural equations and the derived structural equations. Our task now is to find a system of involutive seeds and an involutive ordering. For this system, no index is special and hence we take any ordering in which $i,j,k\dots$ run from $1$ to $n$.

 It is best to arrange the set of invariants themselves systematically in order to arrive at a system of involutive seeds. Since all derived algebraic relations must be taken account of, it is best to arrange the invariants order by order. We will use the following strategy: when using an algebraic relation relating invariants, we prefer to take the invariants having the following property as the independent ones:
\begin{enumerate}
\item They are of the lowest order;
\item Their indices are of decreasing order, if possible.
\end{enumerate}
This ``if possible'' is intentionally vague since it is not an requirement, only a guideline aiming to make the condition \textbf{O1} hold. For the defining relations \eqref{eq:def}, this suggests that we take the following to be independent with respect to this relation:
\[
R_{ijkl} \qquad\text{such that}\qquad i>j,\ k>l.
\]

For the first Bianchi relation \eqref{eq:bianchi1}, we will use a now well-known trick to make our work simpler. We expand the identity
\[
R_{i[jkl]}-R_{j[ikl]}-R_{k[ijl]}+R_{l[ijk]}=0
\]
and use \eqref{eq:def}: this gives
\[
R_{ijkl}=R_{klij},
\]
so in addition, we can require that the independent terms satisfy
\[
R_{ijkl}\qquad\text{such that}\qquad i\ge k.
\]
However, this does not use up all the relations $R_{i[jkl]}=0$. Note that unless all four indices are different, this relation is trivial. Assuming all indices different, and $i>j>k>l$,
\[
R_{ijkl}-R_{ikjl}+R_{iljk}=0
\]
contains two terms for which the second index is greater than the fourth and one term for which the second index is less than the fourth. Hence we can also require that the independent terms satisfy
\[
R_{ijkl}\qquad\text{such that}\qquad j\ge l.
\]
 There are now no more relations on $R_{ijkl}$. Incidentally, this procedure counts the number of algebraically independent terms of $R_{ijkl}$:
\begin{equation}
  \label{eq:1storder}
  i>j,\qquad k>l,\qquad i\ge k,\qquad j\ge l
\end{equation}
giving
\[
2{n\choose 4}+3{n\choose 3}+{n\choose 2}=\frac{n^{2}(n^{2}-1)}{12}
\]
independent terms.

Next we come to the first order invariants, $R_{ijkl;m}$. First we need to take care of the derived relations. This means that we may take $R_{ijkl;m}$ to be independent only if it results from the derivative of an independent fundamental invariant, in other words, we still require \eqref{eq:1storder}. The only additional relation that these invariants are subject to are the second Bianchi relations \eqref{eq:bianchi2}. Using a completely analogous reasoning as the one we used for the Bianchi identity, we require that independent terms satisfies in addition $k\ge m$, in other words, the requirement on the indices is now
\begin{equation}
  \label{eq:2ndorder}
  i>j,\qquad k>l,\qquad i\ge k,\qquad j\ge l,\qquad k\ge m.
\end{equation}
There are no generic algebraic relations since we only have one derivation index.

Having obtained all the algebraic relations we need, we can now verify the condition \textbf{R}: the expansion \eqref{eq:def1stinv} shows that we really have too many invariants for it to fail. For example, for the invariants $R_{ikj\bar k}$, $R_{ni n\bar i}$, $R_{n,n-1,n,n-2}$, the matrix in question is already of maximal rank: it suffices to pay attention to the terms linear in $\omega_{ij}$ in the expansion of $dR_{ikj\bar k}$, the terms linear in $\omega_{i}$ in the expansion of $dR_{ni\bar n\bar i}$, and the term linear in $\omega_{n}$ in the expansion of $dR_{n,n-1,n,n-2}$.

Now we take the independent terms of $R_{ijkl;m}$ as the involutive seeds: they obviously cover all of $R_{ijkl;m}$ since they are obtained by applying all the algebraic relations on $R_{ijkl;m}$, and since all of $R_{ijkl;m}$ are covered, all of $R_{ijkl}$ are as well. The involutive ordering is the ordering that we have been using all along, and that this is an involutive ordering is obvious from the fact that the only requirement on the last index $m$ is $k\ge m$, so if $p<m$, we also have $k\ge p$.

Now, what is the maximal value that can be taken by $m$? Let us try $n$, which is the dimension of the manifold. Since we require
\[
i\ge k\ge m=n,
\]
we have
\[
i=k=m=n, \qquad n>j\ge l,
\]
so the number of involutive seeds with $m=n$ is the same as that of a symmetric bilinear form of dimension $n-1$, namely $n(n-1)/2$. This is the answer that we have been expecting.

Note that we can take the independent invariants at any order to be the involutive seeds. If we do this, besides the derived relations, we need to uphold the generic relations, which means that for our purpose the derivation indices must be non-increasing. Hence, for example, at the third order, the seeds with the maximal last index are $R_{ninj;nnn}$, with
\[
n>i\ge j,
\]
giving the same answer $n(n-1)/2$.

\begin{rmk}
  One is justified to have the feeling that for this problem, it is
  not necessary to formulate the problem in the language of moving
  frames at all. Indeed, let $R_{\mu\nu\rho\lambda}$ be the Riemann
  tensor in the usual language of tensor analysis, then our
  fundamental invariants are just
  \[
  R_{ijkl}=e^{\mu}_{i}e^{\nu}_{j}e^{\rho}_{k}e^{\lambda}_{l}
  R_{\mu\nu\rho\lambda},
  \]
  where $e^{\mu}_{i}$ is the linear transformation that transforms the
  coordinate frame into an orthonormal frame (they are the
  ``vierbein''). And for our purpose, the symmetries of $R_{ijkl}$ and
  those of $R_{\mu\nu\rho\lambda}$ are equivalent since we ignore all non-independent terms.

  It is indeed true that, provided one is very careful, the algorithms
  can be applied to tensors. But using tensor analysis, being careful
  is extremely difficult: with the moving frame, it is simple to
  exhaust all the algebraic relations applicable to the system, and on
  the other hand, using tensor analysis it is very easy to overlook
  one or two relations. Structurally speaking, the
  methods of moving frame are much more rigid than the methods of
  tensor analysis, and there are crazy things that can be done to
  tensors using tensor analysis but are forbidden using moving
  frames. Hence, to ensure the applicability of the algorithm to
  general tensor systems, many more tests and conditions are required,
  which is far more difficult than rewriting the system in the
  language of moving frames.
\hfill\P\end{rmk}

\paragraph{Riemannian geometry with torsion.}

If, instead of writing the structural relations \eqref{eq:rstreq}, we write
\begin{equation}
  \label{eq:rstreqwt}
\left\{
  \begin{aligned}
    d\omega_{i}&=-\omega_{ij}\wedge\omega_{j}+\tfrac{1}{2}T_{ijk}\omega_{j}\wedge\omega_{k},\\
    d\omega_{ij}&=-\omega_{ik}\wedge\omega_{kj}+\tfrac{1}{2}R_{ijkl}\omega_{k}\wedge\omega_{l},
  \end{aligned}
\right.
\end{equation}
we have a Riemannian geometry with torsion $T_{ijk}=-T_{ikj}$. We can carry out the above procedure as before: the algebraic relations for $R_{ijkl}$ are, up to terms in $T_{ijk}$ and $T_{ijk;l}$, which has no effect in our theory, exactly the same as before, and there are no separate Bianchi relations for $T_{ijk}$. Thus, for the involutive seeds, we use $R_{ijkl;m}$ with symmetry as before, and $T_{ijk;l}$. As $T_{ijk;l}$ satisfies no Bianchi relation of its own, there is no restriction on $l$, and for $l=n$ the number of seeds is just the number of independent $T_{ijk}$. As $j$ and $k$ are antisymmetric by the defining relation, the additional contribution is $n^{2}(n-1)/2$.

Actually, since we already know the degree of arbitrariness of Riemannian geometry without torsion, we can reason as follows: the structural equation \eqref{eq:rstreqwt} defines the two one-forms $\omega_{i}$ and $\omega_{ij}$, from which we can define uniquely the two one-forms $\omega_{i}$ and $\omega'_{ij}$, such that these two forms satisfy \eqref{eq:rstreq}, by setting
\[
\omega_{ij}'=\omega_{ij}+(-T_{ikj}+T_{jki}+T_{kji})\omega_{k}.
\]
Hence the system \eqref{eq:rstreqwt} is completely equivalent to the system \eqref{eq:rstreq} together with the additional function $T_{ijk}$ with the relations
\[
T_{ijk}=-T_{ikj},
\]
from which we can easily deduce the additional contribution to the degree of arbitrariness.

\paragraph{Ricci-flat spacetime. Einstein equations.}

Now let us return to the theory without torsion, and focus on Riemannian spaces that are Ricci-flat. Such spaces are defined by the constraint
\[
R_{ij}\equiv\sum_{k}R_{kikj}=0.
\]
Note that this constraint is equivalent to the vacuum Einstein equations 
\[
R_{ij}-\tfrac{1}{2}\delta_{ij}R=0,\qquad R\equiv \sum_{i}R_{ii},
\]
and hence Ricci-flat spacetimes are just those satisfying Einstein equations in vacuum (and without cosmological constant). ({We have pretended that Einstein's equations are defined using a positive-definite metric. The construction and analysis, done properly using metric with the correct signature, is essentially identical to what we do here.}) Remark that these constraints, when written in terms of the metric, are differential equations, but for us they involve no differentiation at all.

For this system, we can make the following changes to our calculation for the general Riemannian case to obtain the new degree of arbitrariness: for $i,j<n$, the equations we have added are
\[
R_{1i1j}+R_{2i2j}+\dots+R_{ni nj}=0,
\]
so now, by the condition \textbf{O1}, $R_{ni nj}$ for $i,j<n$ are no longer considered independent terms. For $i= n$, $j<n-1$, we have
\[
R_{1n1j}+R_{2n2j}+\dots+R_{n-1,n,n-1,j}=0,
\]
so $R_{n,n-1,n-1,j}$ for $j<n-1$ are no longer considered independent. For $i=n$, $j=n-1$,
\[
R_{1,n,1,n-1}+R_{2,n,2,n-1}+\dots+R_{n-2,n,n-2,n-1}=0,
\]
so $R_{n,n-2,n-1,n-2}$ is no longer considered independent. Finally, for $i=n$, $j=n$,
\[
R_{n1n1}+R_{n2n2}+\dots+R_{n,n-1,n,n-1}=0.
\]
However, all of these terms have already been declared dependent in the case of $i,j<n$. Hence we need to substitute them with independent terms. After this substitution, we see that $R_{n-1,n-2,n-1,n-2}$ is no longer independent.

In summary, the independent (normal) components of the Riemann tensor are now $R_{ijkl}$ for which
\[
i>j,\qquad k>l,\qquad i\ge k,\qquad j\ge l,
\]
from which we exclude the following
\[
R_{n i n j},\qquad R_{n,n-1,n-1,i},\qquad R_{n,n-2,n-1,n-2},\qquad R_{n-1,n-2,n-1,n-2}.
\]

We also need to take the equations $dR_{ij}=0$ into account. The derived equations only tell us that {independent derived invariants comes from derivation of independent fundamental invariants.}

Now let us apply our algorithms. As in the unconstrained case, the verification of condition \textbf{R} is easy. The largest possible last index for the seeds is now $n-1$, and we have the following possibilities:
\[
R_{n-1,i,n-1,j;n-1},\qquad R_{n,i,n-1,j;n-1}.
\]
The first possibility gives
\[
\frac{(n-1)(n-2)}{2}-1
\]
terms (we need to exclude $R_{n-1,n-2,n-1,n-2;n-1}$), and the second possibility also gives
\[
\frac{(n-1)(n-2)}{2}-1
\]
terms (we must have $i<n-1$, and exclude $R_{n,n-2,n-1,n-2}$). Hence the degree of arbitrariness is now
\[
n(n-3)
\]
at dimension $n-1$. 

For the general Einstein equation
\[
R_{ij}-\tfrac{1}{2}\delta_{ij}R+\delta_{ij}\Lambda=T_{ij},
\]
the degree of arbitrariness in the general case is the same as in the vacuum case, as can be verified by a similar but slightly more complicated calculation, by using the expression for the Ricci scalar $R$.

It should be noted that $n(n-3)$ is always an even number, and this shows that the Einstein equations are well-behaving dynamical equations suitable for describing particles: the number $n(n-3)/2$ is of course just the degree of \emph{freedom} of the graviton, which when $n-4$ yields the familiar answer $2$.

Let us now briefly consider Einstein--Cartan theory, in which we have Riemannian geometry with torsion. In addition to the usual Einstein equations, we couple the torsion to spin density directly:
\[
T_{ijk}=S_{ijk}
\]
so that all degree of arbitrariness generated by $T_{ijk}$ are killed. Thus, the theory has exactly the same degree of arbitrariness as the usual Einstein theory. Physically, we say that ``spin density does not propagate'': there are no ``spin waves''.

\begin{rmk}
  As treated here, the direction of evolution of Einstein's equations can be taken to be any direction, i.e., any submanifold can be used as a Cauchy surface. But this is only due to the fact that we are formulating the theory as if the metric is positive-definite: in this case the system admits no characteristic directions. If we use the metric of the correct signature, or what amounts to the same thing, if we write the structural equations \eqref{eq:rstreq} as the structural equations for the group $SO(n-1,1)\times M$ instead of $SO(n)\times M$, which in practice just adds a few minus signs here and there, one important change to the theory is that the theory now admits characteristic directions, which are just the null directions. As we have discussed, the Cauchy surface must not contain characteristics, and hence the signature forces the time direction to be the direction of evolution.\hfill\P
\end{rmk}

\begin{rmk}
  If we compare the theory before and after adding the Einstein equations as constraints, we see that the dimension at which the degree of arbitrariness occurs is exactly one lower than the unconstrained theory. This property, together with the property mentioned above about the time direction, are the requirements for any physically meaningful equations of motion: the theory must be neither ``over-determined'' nor ``under-determined'' by a suitable Cauchy data.
\hfill\P
\end{rmk}

\begin{hnt}\label{ba}
  For $n\ge 4$, the degree of arbitrariness $n(n-3)$ we obtained for
  Einstein's equations is greater than the number of degree of arbitrariness
  for a $n-1$ dimensional Riemannian space, but less than the degree
  of freedom of a $n-1$ dimensional Riemannian space plus a symmetric
  bilinear form (For $n=3$, the degree of arbitrariness by our formula is
  zero: we can see this even without calculating since $R_{ij}=0$
  implies $R_{ijkl}=0$ when $n=3$). Hence we may speculate that the
  degree of arbitrariness that we have obtained decomposes into the metric
  on the $n-1$ dimensional ``section'' and the second fundamental form
  on this section, subject to certain constraints. This suggests that
  we can try to use a metric on a hypersurface and a second
  fundamental form subject to certain constraints as the Cauchy data
  for the Einstein equations. That such a specification actually works
  is shown by Choquet-Bruhat \cite{mdm-o} in her study of the Cauchy problem
  of Einstein's equations: for a modern account, see 
  \cite{mdmb}. However, since the second fundamental form is subject to the
  Gauss-Codazzi equation, and in Choquet-Bruhat's approach there are
  further constraints on them (the momentum constraint and the
  Hamiltonian constraint), it is
  impossible to derive the degree of arbitrariness in this way without a
  large amount of further work. On the other hand, before the work of
  Choquet-Bruhat, Cartan has already derived the correct degree of
  freedom of the vacuum Einstein equation $n(n-3)$ at dimension $n-1$
  by a rather ad-hoc reasoning, which is complicated by the
  arbitrariness of coordinates, in his correspondence with Einstein: see \cite{cartan-einstein}, letter \textsc{xxii}.
\hfill\P\end{hnt}

\subsection{Gauge theories and Yang--Mills equations}

Here we want to find the degree of arbitrariness of a classical gauge theory over a Riemannian geometry. A gauge theory on a Riemannian manifold is usually specified by writing down some gauge potentials $A_{ai}$, for which $a$ is a group index (omitted if the group is one-dimensional) and $i$ is a spacetime index, and $A_{ai}$ depends on the spacetime coordinates only. This requires $pn$ functions to define, where $p$ is the dimension of the group and $n$ is the dimension of spacetime. However, there is also the ``gauge invariance'' of $A_{ai}$ that must be taken into account: thus, the true degree of arbitrariness is, by this intuitive argument, $p(n-1)$.

Let us calculate this number by our algorithm. In the formulation of moving frames, a gauge theory is constructed as follows: we have the structural equations for the Riemannian space \eqref{eq:rstreq}, and we couple a Lie group $G$ to it. Let us assume that the Lie group has Maurer-Cartan structural equations
\[
d\alpha_{a}=-C_{abc}\alpha_{b}\wedge\alpha_{c},
\]
where $\alpha_{a}$ are the Maurer-Cartan forms and $C_{abc}$ are the structure \emph{constants} for the Lie group. We form the product space of the Riemannian principal bundle and the Lie group, and change the structural equation to
\[
d\gamma_{a}=-C_{abc}\gamma_{b}\wedge\gamma_{c}+\tfrac{1}{2}F_{aij}\omega_{i}\wedge\omega_{j}.
\]
The defining relations for $F_{aij}$ are as follows: for the index $a$, the symmetry is the same as that of the form $\alpha_{a}$. For the index $i,j$, we have $F_{aij}=-F_{aji}$. There is also an additional relation occurring at first order, namely
\[
dF_{aij}=F_{aij;k}\omega_{k}+G_{bij}\gamma_{b}+(\text{terms in $\omega_{ij}$}),
\]
where $G_{bij}$ are functions of $F_{aij}$, the exact form depending on the group. For example, for the group $SO(p)$, the additional one-forms are $\gamma_{ab}$ with $a,b$ antisymmetric,
\[
dF_{abij}=F_{abij;k}\omega_{k}-F_{cbij}\gamma_{ac}-F_{acij}\gamma_{bc}-F_{abkj}\omega_{ik}-F_{abik}\omega_{jk}.
\]
Basically, the form of these relations just means that the group directions are not dynamical, as differentiations in these directions do not generate new independent invariants.

The Bianchi relations for $F_{aij}$ is
\[
F_{a[ij;k]}=0,
\]
and the rest of the relations are obvious. Thus, we can take the independent terms of $F_{aij}$ to be those that have $i>j$, and using the Bianchi relation, take the independent terms of $F_{aij;k}$ to be those that have
\[
i>j,\qquad i\ge k.
\]
Then we take the set of involutive seeds be the independent $R_{ijkl;m}$ as before, and also the independent $F_{aij;k}$. The involutive order is the order that all indices $a,b,c\dots$ are considered greater than $i,j,k\dots$. The number of terms $F_{ani;n}$ is thus
\[
p(n-1),
\]
where $p$ is the dimension of the Lie group, since the index $a$ in $F_{aij}$ is a group index. The condition \textbf{R} can be easily verified as before.
 Hence this is the additional contribution to the degree of arbitrariness, occurring at dimension $n$, as the not-so-precise argument at the beginning of this section shows. For example, for $SO(2)$ gauge theory (electromagnetism) in dimension $3+1$, this number is simply $3$.

\paragraph{Yang--Mills equations.}

We now study the degree of arbitrariness of adding to the above system the classical Yang--Mills equation, which in our notation, reads
\[
\sum_{i}F_{aij;i}=\text{source terms},
\]
which has a total of $pn(n-1)/2$ equations: more than the degree of arbitrariness of the original system. For the equations of motion, when $j<n$, the constraints are just that
\[
F_{anj;n}
\]
are no longer independent by condition \textbf{O1}. When $j=n$, we have (omitting the group indices)
\[
F_{1n;1}+F_{2n;2}+\dots+F_{n-1,n;n-1}=\cdots,
\]
so $F_{n,n-1;n-1}$ is no longer independent. Under Yang--Mills equation for the gauge fields \emph{and} Einstein's equations for the gravitational field, the additional contribution to the degree of arbitrariness from the gauge fields is therefore the number of remaining normal terms of $F_{aij;n-1}$, which are
\[
F_{a,n,j;n-1}\quad(j=1,\dots,n-2),\qquad F_{a,n-1,j;n-1}\quad(j=1,\dots,n-2),
\]
giving the degree of arbitrariness
\[
2(n-2)r.
\]
occurring at dimension $n-1$.
This gives the number of fields we must specify on a hypersurface to have a well-defined Cauchy problem for Yang--Mills equations coupled to Einstein gravity, or in flat spacetime. Note that it is essential to impose the Einstein equation as well, otherwise the $2(n-2)r$ degree of arbitrariness we get here at dimension $n-1$ will be eclipsed by the degree of arbitrariness of the Riemannian geometry at dimension $n$: more on this in the next section. Again, an essential feature is that when going up one dimension, we require \emph{two} additional copies of the Lie algebra: this corresponds to \emph{one} degree of freedom for the boson. When $r=1$ and $n=4$, we see that the photon has $2$ degrees of freedom.

\begin{hnt}
  For specific dimensions and specific groups, Estabrooks \cite{edsym} has
  set up explicit exterior differential systems for the Yang--Mills
  equations using coordinates, and, with the help of computer algebra programs, calculates
  the Cartan characters for the systems (which, as we shall see in the
  proof of our algorithm, is intimately related to the degree of
  freedom). For example, for $SU(2)$ Yang--Mills equations of
  dimensions $3,4,5,6$, the last non-vanishing Cartan characters occur
  at dimensions $2,3,4,5$, and are $9,15,21,27$. However, since
  Estabrooks used coordinates, the gauge degree of arbitrariness ($3$ for
  $SU(2)$ Yang--Mills theory) is still present. If we subtract it from
  his answers, we get $6,12,18,24$, which is just our answer. The Cartan characters obtained for Maxwell theory for dimensions $3,4,5,6$ by Estabrooks can also be shown to be in complete agreement with our result by analogous reasoning. Of
  course, our algorithm is so simple so that it is unnecessary to
  resort to computers for the calculations, we not need to reason with
  the gauge degrees of arbitrariness since no coordinate is used, and our
  result does not depend on the dimension nor on the geometry of the underlying space (Estabrooks considered only theories set up in flat spaces).\hfill\P
\end{hnt}

\subsection{Scalar field theory on Riemannian manifold}
We have checked that our algorithm gives the correct answers for gravitons and gauge bosons. Let us now very briefly check the case of scalar fields.
In our approach, adding a scalar field corresponds to having a free function to a Riemannian manifold. From our discussion about additional functions, it is obvious that at the kinematical level, the extra degree of arbitrariness is $1$: denoting the field be the scalar function $f$, this degree of arbitrariness comes from $f_{;n}$. Now we study what happens when we specify the equations of motion for the scalar field, i.e., its dynamics.

 We add the Klein--Gordon equation, which is, in moving frames, the single equation
\[
\sum_{i=1}^{n}f_{;ii}=m^2 f.
\]
Assuming that we have already killed all degrees of arbitrariness coming from the Riemannian metric. For any system of involutive seeds and ordering, $f_{;nn}$ is no longer considered a seed. Hence, by the generic relations, the degree of arbitrariness on a Cauchy surface comes from the two terms
\[
f_{;n,n-1},\qquad f_{;n-1,n-1}
\]
and thus the degree of arbitrariness of a scalar field under Klein--Gordon equation is exactly $2$, independent of the dimension of the manifold, occurring at dimension $n-1$. Of course, this translates to a degree of freedom of scalar particles $\tfrac{2}{2}=1$.
\subsection{Newtonian rigid motion}

Of course, the point of this algorithm is not for studying the degrees of freedom of elementary particles, which are already known. Thus let us now consider constrained systems, the first example of which being the familiar Newtonian rigid motion.

Intuition tells us that a Newtonian rigid motion has six degree of arbitrariness. As a Newtonian rigid motion is locally just a rigid flow (i.e., shear-free and expansion-free flow), this is also the degree of arbitrariness of such a flow in fluid mechanics, but note that in our language, this degree of arbitrariness occurs at dimension $1$, even in the fluid description. Below we shall derive this rigorously using our algorithm. This example also shows a few pitfalls when applying our procedure.

Newtonian motion is formulated on a space with a Galilean connection defined on it. The connection matrix can be written
\[
\begin{pmatrix}
  0&0&0\\
  \tau&0&0\\
  \theta_{i}&\omega_{i}&\omega_{ij}
\end{pmatrix}
\]
with the additional constraints
\[
d\tau=0
\]
which guarantees the existence of absolute time,
\[
d\omega_{ij}=-\omega_{ik}\wedge\omega_{kj}
\]
which guarantees the existence of flat space, and
\[
(d\omega_{i}+\omega_{ij}\wedge\omega_{j})\wedge\theta_{i}=0
\]
which guarantees the absence of velocity-dependent gravitational effect. For how these equations are derived, see \cite{cartan-newtonian}, or the English translation \cite{cartan-n-t}. In the usual coordinates, this matrix is of the form
\[
\begin{pmatrix}
  0&0&0\\
  dt&0&0\\
  dx_{i}&G_{i}dt&0
\end{pmatrix},
\]
but we will not use any of such coordinates. This moving frame can be adapted to the {rigid motion} of our interest if we align the co-frame $\tau$  with the direction of motion. In the language of fluid mechanics, the flowlines, characterised by $\tau$, drags the spatial structure along while keeping them unchanged. The structural equation of a rigid motion is 
\[
\left\{
  \begin{aligned}
    d\tau&=0,\\
    d\theta_{i}&=-\omega_{i}\wedge\tau-\omega_{ij}\wedge\theta_{j},\\
    d\omega_{i}&=-\omega_{ij}\wedge\omega_{j}+\Gamma_{ij}\theta_{j}\wedge\tau+\tfrac{1}{2}\Gamma_{ijk}\theta_{j}\wedge\theta_{k},\\
    d\omega_{ij}&=-\omega_{ik}\wedge\omega_{kj},
  \end{aligned}
\right.
\]
which are just the structural equations for Newtonian spacetime (the one-form $\tau$ implicitly has the index $0$), \emph{together} with the decomposition
\[
\omega_{i}=K_{i}\tau+M_{ij}\theta_{j}.
\]
For more details, see \cite{hu1}.
Thus, properly speaking, the co-frame is really formed by $\tau$, $\theta_{j}$ and $\omega_{ij}$: the $\omega_{i}$ are really just convenient one-forms to work with.
Roughly speaking, $K_{i}$ signifies the acceleration and $M_{ij}$ the rotation.
 The fundamental invariants are hence $\Gamma_{ij}$ and $\Gamma_{ijk}$, with a defining relation
\[
\Gamma_{ijk}=-\Gamma_{ikj},
\]
together with $K_{i}$ and $M_{ij}$, with $M_{ij}=-M_{ji}$. The condition of the absence of velocity-dependent gravitational effect implies
\[
\Gamma_{[ijk]}=0,\qquad \Gamma_{[ij]}=0.
\] 
Here some of the Bianchi identities are obtained by expanding
\[
d\omega_{i}=d(K_{i}\tau+M_{ij}\theta_{j})
\]
using the structural equations, which gives
\begin{equation}
  \label{eq:offd1}
  M_{ij;k}=0,\qquad K_{i;j}=M_{ij;0}+M_{ik}M_{kj}.
\end{equation}
The other Bianchi relations are easily derived to be
\[
\Gamma_{ijk}=0,\qquad \Gamma_{i[j;k]}=0,
\]
but as $\Gamma_{[ij]}=0$, $\Gamma_{ij;k}$ is totally symmetric in all three indices. It is easy to check that we again have an abundance of invariants, so that condition \textbf{R} is satisfied.

It is attempting to take the set
\[
K_{i;j},\qquad K_{i;0},\qquad \Gamma_{ij;0}\quad(i\ge j)\qquad \Gamma_{ij;k}\quad(i\ge j\ge k)
\]
as the set of involutive seeds. However, this does not work: the second relation in \eqref{eq:offd1} means that the condition \textbf{O1} fails for $K_{i;j}$ if the involutive ordering has $0$ as the smallest index, and the condition \textbf{O2} fails for $\Gamma_{ij;k}$ and $\Gamma_{ij;0}$ if $0$ is not taken as the smallest index. Instead, we must take the system of involutive seeds to be
\[
M_{ij;0},\qquad K_{i;0},\qquad \Gamma_{ij;0}\quad(i\ge j)\qquad \Gamma_{ij;k}\quad(i\ge j\ge k)
\]
and specify the involutive ordering as $0<i,j,k,\dots$, the ordering among $i,j,k\dots$ being immaterial. We see that the degree of arbitrariness is $1$, given by the term $\Gamma_{n-1,n-1;n-1}$, occurring at dimension $n$.

But actually this system is \emph{not} exactly what we really want to use if we want to calculate the degree of arbitrariness of Newtonian rigid motion: the presence of the derivatives of $\Gamma_{ij}$ means that the degree of arbitrariness of the gravitational field is still present. The degree of arbitrariness of the gravitational field is also $1$ at dimension $n$: we can calculate this separately but it is most easily seen if in the above system of involutive seeds we simply remove $M_{ij;0}$ and $K_{i;0}$. We say that the degree of arbitrariness of the rigid motion is completely eclipsed by the degree of arbitrariness of the gravitational field.

Hence let us now assume that this gravitational degree of arbitrariness is removed by specifying completely the gravitational field. Requiring that the space is Galilean has this effect, as well as using Newton's equation \emph{together with} the appropriate boundary conditions.

This done, all invariants involving $\Gamma_{ij}$ and its derivatives are removed from our system. In order to satisfy condition \textbf{O1}, we need to take the involutive seeds as
\[
M_{ij;0},\qquad K_{i;0}
\]
with the $0$ as the smallest index. The condition \textbf{R} is satisfied for a general system: it suffices to use $M_{ij}$ to solve for $\omega_{ij}$, $K_{i}$ to solve for $\theta_{i}$, and $K_{i;0}$ to solve for $\tau$. 
Hence, the degree of arbitrariness is
\[
\frac{n(n-1)}{2}
\]
occurring at dimension $1$. For the usual dimension $n=3+1$, this is $6$. The dimension of the degree of arbitrariness, together with the form of the involutive seeds we use, means that we can specify arbitrarily the time derivative of the rotation and acceleration of the rigid body at any given time. The directions of evolution are all of the spatial directions: thus the constraints we have added do not have interpretations of proper equations of motion.

\begin{rmk}
With this framework, deriving the degree of freedom of classical gravitational field is trivial. Indeed, Poissons' equation is simply
\[
\sum_{i=0}^{n-1}\Gamma_{ii}=\text{source terms},
\]
and consequently the contributions to the degrees of arbitrariness is $2$, giving the degree of freedom of the classical gravitational field $1$, similar to the scalar field case.
\hfill\P
\end{rmk}

\subsection{Relativistic rigid flow}

It is perhaps an overkill to use our algorithm to study Newtonian rigid flow. But when it comes to relativity, the situation is vastly different. Shortly after the birth of the special theory of relativity, Born \cite{born1909} defined the concept of rigid flow, as the light-like flow that drags the orthogonal projection of the spacetime metric onto itself. In other words, for a rigid flow, the infinitesimal distance between neighbouring pairs of particles measured orthogonal to the worldline of either remains constant. Such a definition remains valid when we are in a general Riemannian spacetime.

 Whereas rigid flow in classical physics are simple, thanks to the absolute splitting of space and time, in relativity the situation is very different due to the close interplay between space and time. Thus, there is no intuitively obvious answer to the following two questions:
\begin{enumerate}
\item What is the degree of arbitrariness, including the degree of arbitrariness of the underlying space, of a relativistic rigid flow?
\item What is the degree of arbitrariness of a relativistic rigid flow in a \emph{specified} spacetime?
\end{enumerate}

In the moving frame formulation, a rigid flow in a Riemannian spacetime is completely specified by the moving frame
\[
\left\{
  \begin{aligned}
    d\omega_{i}&=-\pi_{ij}\wedge\omega_{j},\\
    d\omega_{0}&=-K_{i}\omega_{0}\wedge\omega_{i}-M_{ij}\omega_{i}\wedge\omega_{j},\\
    d\pi_{ij}&=-\pi_{ik}\wedge\pi_{kj}+\tfrac{1}{2}S_{ijkl}\omega_{k}\wedge\omega_{l},
  \end{aligned}
\right.
\]
where $\omega_{0}$ is aligned with the flowline, $\omega_{i}$ the orthogonal directions, and the connection for the Riemannian spacetime is given by
\begin{equation}
  \label{eq:rfconn}
  \omega_{0i}=-\omega_{i0}=K_{i}\omega_{0}-M_{ij}\omega_{j},\qquad \omega_{ij}=\pi_{ij}+M_{ij}\omega_{0}.
\end{equation}
The derivation is slightly non-trivial: see \cite{hu1}. As in the Newtonian case, $K_{i}$ and $M_{ij}$ has the physical interpretations of acceleration and rotation (usually called vorticity in fluid) respectively. $S_{ijkl}$ is the Riemann tensor for the ``reduced space'', analogous to the Riemann tensor fixed to the moving body in classical rigid motion.

The fundamental invariants are thus $K_{i}$, $M_{ij}$ and $S_{ijkl}$. $S_{ijkl}$ has the same defining and Bianchi relations as the usual Riemann tensor. $M_{ij}$ has defining relation
\[
M_{ij}=-M_{ji}
\]
and we can derive the following additional Bianchi relations
\begin{equation}
  \label{eq:addbian}
    M_{[ij;k]}=3M_{[ij}K_{k]},\qquad
    M_{ij;0}=-K_{[i;j]},\qquad S_{ijkl;0}=0.
\end{equation}
Again we have abundance of invariants so that the condition \textbf{R} is satisfied.
We can thus take the involutive seeds to be
\[
M_{ij;k}\quad(i>j,\ i\ge k)\qquad K_{i;0}\qquad K_{i;j}\qquad S_{ijkl;m}\quad (i>j,\ k>l,\ i\ge k,\ j\ge l,\ k\ge m),
\]
and the involutive ordering is the ordering such that $0$ is the \emph{greatest} index: this is different from the classical case, where we are forced to take the time index to be the \emph{smallest}. The degree of arbitrariness is thus the number of seeds of the form $K_{i;0}$, which means that the system has degree of arbitrariness
\[
n-1
\]
occurring at dimension $n$.

Remark that we have \emph{not} yet specified the geometry of the Riemannian space for which the flow takes place. In the classical case, we see that the degree of arbitrariness of the rigid flow is completely eclipsed by the degree of arbitrariness of the underlying geometry, but here, a general Riemannian geometry has degree of arbitrariness $n(n-1)/2$, which for $n>3$ is greater than the degree of arbitrariness $n-1$ that we have obtained. This means that \emph{not all Riemannian spaces with $n>3$ admit rigid flows} ({Or, more precisely, \emph{almost all} Riemannian spaces with $n>3$ \emph{do not} admit rigid flows. For $n=2$, the situation is different, and there always exists rigid flows for any Riemannian space: see \cite{hu1}).

What are these $n-1$ degrees of arbitrariness? We see that we can freely specify the time derivative $K_{i;0}$ everywhere. By taking $K_{i}$ as known functions and use our algorithm again, we can show that it is also consistent to freely specify $K_{i}$. However, it is in general not consistent to specify the geometry $S_{ijkl}$ of the reduced space independently in space: in relativity, we cannot start with a ``shape'' that we want to move around in spacetime and drag it along to obtain a rigid flow, in constrast with the Newtonian case.

We now come to the second question, namely the degree of arbitrariness in a specified spacetime. Using \eqref{eq:rfconn}, the invariants $R_{\mu\nu\rho\lambda}$ of the total space is related to our invariants as follows
\begin{equation}
  \label{eq:bornstreqns}
  \left\{
  \begin{aligned}
    R_{ijkl}&=S_{ijkl}+M_{il}M_{jk}-M_{ik}M_{jl}-2M_{ij}M_{kl},\\
    R_{ijk0}&=M_{ij;k}-M_{jk}K_{i}+M_{ik}K_{j}+M_{ij}K_{k},\\
    R_{0i0j}&=M_{ik}M_{jk}-K_{(i;j)}-K_{i}K_{j}.
  \end{aligned}
\right.
\end{equation}
To completely specify a spacetime means simply to take $R_{\mu\nu\rho\lambda}$, as well as all of their derivatives, as known functions. The first relation can be used to express $S_{ijkl}$ completely in terms of $M_{ij}$ and $R_{ijkl}$, the second can be used to express $M_{ij;k}$ completely in terms of $R_{ijk0}$, $M_{ij}$ and $K_{i}$ (the reader can check that this is really the case, by verifying that the symmetries of the equation does not force us to miss any terms), and the third equation expresses $K_{(i;j)}$ completely in terms of $R_{0i0j}$, $K_{i}$ and $K_{j}$. By considering $R_{ijkl;0}$, we see that \emph{in the general case} $M_{ij;0}$ is completely specified in terms of $R_{ijkl;0}$: in more details, the first equation implies
\[
R_{ij\bar i\bar j}=S_{ij\bar i\bar j}-3M_{ij}M_{\bar i\bar j}
\]
where a bar over $i$ means that there is no summation between $i$ and $\bar i$, but otherwise $i=\bar i$. This implies
\[
M_{ij;0}M_{\bar i\bar j}=-\tfrac{1}{6}R_{ij\bar i\bar j;0}
\]
and if $M_{ij}\neq 0$ for all $i,j$, which is what we mean by the general case, $M_{ij;0}$ is completely determined by $R_{ij\bar i\bar j;0}$, which by our assumption is a known function. Now we see that $S_{ijkl}$, $M_{ij;k}$, $M_{ij;0}$, $K_{(i;j)}$ are all completely determined. By \eqref{eq:addbian}, $K_{[i;j]}$ is as well. Differentiating the second equation of \eqref{eq:bornstreqns} and taking those linear in $\omega_{0}$ shows that \emph{in the general case}, $K_{i;0}$ is completely determined. By our definition in section \ref{sec:invseeds}, the invariants are covered by the empty set, and hence this system has no degree of arbitrariness at all. 
This means that the system either has solutions depending on constants, or has no solutions at all. This is in accordance with our remark earlier that, since the degree of arbitrariness of the system without specifying the geometry is less than the degree of arbitrariness of a general Riemannian space, not all configurations admit solutions. In fact, if the space is homogeneous, i.e., $R_{\mu\nu\rho\lambda}$ is a single constant, then a generalisation of the Herglotz--Noether theorem applies, and all solutions are rotating Killing vectors in the space.

What about the non-general case? For simplicity, we study only the case where the total space is homogeneous. We need to consider only the case where $M_{ij}=0$ for all $i,j$, since otherwise by using the right action of the principal bundle we can transform the system locally into an equivalent one for which $M_{ij}\neq 0$ for all $i,j$. Then, the equations \eqref{eq:bornstreqns} become
\[
\left\{
  \begin{aligned}
    R_{ijkl}&=S_{ijkl},\\
    R_{ijk0}&=0,\\
    R_{0i0j}&=-K_{(i;j)}-K_{i}K_{j}.
  \end{aligned}
\right.
\]
Immediately we see from the second equation above that, for homogeneous space, unless the space has vanishing curvature, i.e., Minkowski space, there is no singular solution. Hence assume that we are in Minkowski space. Then
\[
S_{ijkl}=0,\qquad K_{(i;j)}=-K_{i}K_{j},
\]
which together with $K_{[i;j]}=0$, forces us to take the involutive seeds as
\[
K_{i;0}.
\]
Let us now come to the verification of condition \textbf{R}. In the general case, we use $K_{i}$ to solve for $\omega_{i}$, $K_{i;0}$ to solve for $\omega_{0}$. As for $\omega_{ij}$, we use the series of invariants $K_{i;00},K_{i;000},K_{i;0000},\dots$: this is possible, since all of these transform under $SO(n-1)$ and are independent.

The degree of arbitrariness is thus $n-1$, occurring at dimension $1$. Intuitively, this is the motion of hyperplanes in Minkowski spacetime, the hyperplane always being orthogonal to the worldline of any one of its point. The $n-1$ degree of arbitrariness is the acceleration of any one of its point as a function of time.

\begin{hnt}
  The reference \cite{Giulini:2006p66} has a good summary of the history and proof of the Herglotz--Noether theorem mentioned above, except that it missed out the contributions of Estabrook and Wahlquist \cite{Estabrook:1964p2608,Wahlquist:1966p2548,Wahlquist:1967p2556}, which generalised the theorem to conformally flat spacetime. The Herglotz--Noether theorem and its generalisation by Estabrook and Wahlquist are both resticted to $3+1$ dimensions since the problem was phrased in a form that is unsuitable for calculations in arbitrary dimensions: this is especially the case for the generalisation, since the dyadic method used by Estabrook and Wahlquist do not apply in higher dimensions at all.  The present author, on the other hand, has generalised them further to all dimensions and all conformally flat spaces, by formulating Born-rigid flows as \emph{structure-preserving submersions}, of which the present example is a very short preview.

In fact, structure-preserving submersions, when the ``structure'' in question is the Riemannian metric, as it is the case here, is just the familiar problems of Riemannian submersions (the reference \cite{Alvarez:2007p4209} has a derivation of the structural equations, see also the references contained therein for further references on Riemannian submersions). But we can also have more exotic structures that are preserved: for example, if a Weyl structure is preserved in a Riemannian spacetime, in the codimension $1$ case the equations describe precisely flows with vanishing shear but possibly non-zero expansion \cite{hu1}. Needless to say, the present algorithm is very useful in such studies since in all cases the degree of arbitrariness can be very easily obtained.
\hfill\P
\end{hnt}

\section{Proof of the algorithm}
\label{sec:pr}

Now we prove our algorithm. The proof consists of three parts:
\begin{enumerate}
\item Construct an exterior differential system for all the invariants of the system that must be satisfied by any solution of the system;
\item Calculate the degree of arbitrariness (the number of free functions) of the general solution to this exterior differential system;
\item Prove that the solution that we obtained is compatible with the \emph{original} exterior differential system, i.e., the structural equations of the moving frame, hence the degree of arbitrariness we obtained really is the degree of arbitrariness of the system we are interested in.
\end{enumerate}

But before that, we need to state the mathematical theorems that we will use. It turns out that, for stating these theorems, we need to define a long train of concepts that are not familiar to the usual physicist. We put these constructions and statements of theorems in appendix \ref{sec:ckip}, and from now on we will use the constructions and results without further explanation. See \cite{cartan-eds, AndrewIvey:2003p7876, LBryant:1991p8950} for more in-depth exposition on the theory that we use.

\subsection{The exterior differential system for the invariants}

First we shall assume that the condition \textbf{R} is satisfied, and there are no additional functions added to the system.  We now consider all the algebraically independent invariants to be independent variables. A subspace formed by a subset of these invariants will be the space on which we work.

Since two functions $f$, $g$, are \emph{functionally independent} if and only if their differentials $df$, $dg$ are \emph{linearly independent}, for systems satisfying the condition \textbf{R}, we can find a full set of independent invariants such that the co-frame $\omega_{\mu}$ is solvable in terms of functions of these invariants. Thus, our original structural equations are now considered differential systems whose variables are the invariants: we do not need any coordinates. ({More precisely, in Cartan's theory of equivalence \cite{AndrewIvey:2003p7876,cartanequ,JOlver:1995p9015,meq}, it is the functional inter-dependence of the invariants that specify completely the solution of the system, not the explict analytic form taken by these invariants.})

Now for any invariant $I_{I;J}$, we have the expansion
\[
dI_{I;J}=I_{I;Jk}\omega_{k}+C_{I;J\alpha}\omega_{\alpha}
\]
which by our assumption, only $I_{I;Jk}$ may contain new algebraically independent invariants. We will take such equations for a certain set of invariants to be the differential system for the invariants, taking care to constrain them by all the Bianchi relations of the systems. The forms $\omega_{k}$, $\omega_{\alpha}$ are now considered to be nothing more than shorthands for some linear combinations of the differentials of the invariants. This is an exterior differential system with independence condition: the independent one-forms are exactly the one-forms $\omega_{i}$, $\omega_{\alpha}$, which gives an implicit independence condition for the invariants themselves. For the moment we will ignore the original structural equations.

Our differential system is closed by adding the equations obtained by exterior differentiation:
\begin{equation}
  \label{eq:neweds}
  0=d^{2}I_{I;J}=d(I_{I;Jk}\omega_{k}+C_{I;J\alpha}\omega_{\alpha}).
\end{equation}
Observe that, if the invariants $I_{I;Jkl}$ for all $l$ are included in our sets of invariants, then such equations are identically satisfied: parts of the algebraic relations satisfied by $I_{I;Jkl}$ are exactly those that ensures this equation holds. Thus, for the two-form equations in our differential system, we only need to consider those coming from the differentiation of one-form equations consisting of the most number of derivation indices.

Expanding \eqref{eq:neweds} and using the equations of the differential systems themselves, we get
\begin{equation}
  \label{eq:2feqs}
  0=dI_{I;Jk}\wedge\omega_{k}+\cdots
\end{equation}
where the dots denote two-forms formed with the independent one-forms. By the previous remark, such equations are non-vacuous only in the cases where derived invariants of $I_{I;Jk}$ are not included in our system, in which case $dI_{I;Jk}$ are now independent one-forms, and hence they are the only two-form equations in our differential system. 

Observe that our definition for a set of involutive seeds is tailored such that, for the involutive seeds and all the invariants covered by them, a set of independent and spanning one-forms $dI_{I;Jk}$ can be taken exactly as those of the derivatives of the involutive seeds: the condition \textbf{I1} ensures that the system includes all the invariants that must be included, and \textbf{I2} ensures independence of the seeds. The case where the set of involutive seeds is empty occurs if and only if when including enough differential invariants into our system, the system is Frobenius-integrable (or inconsistent), and hence contains no degree of arbitrariness by Frobenius theorem.

\subsection{The degree of arbitrariness of the system of invariants}

Now we apply the Cartan--K\"ahler theory to our differential system. We will use version 2 of the involutivity theorem, and hence what we are interested in is the reduced characters without taking into account of the maximal rank condition. As the involutivity test does not need $s'_{0}$, and all the Cartan characters except $s'_{0}$ depend only on equations formed with two-forms and higher order forms, we only need to focus on \eqref{eq:2feqs}. The reduced characters in this case are calculated by feeding the two-forms an independent vector direction so that it becomes an one-form equation, and ignore all the independent forms. Thus, due to the form of \eqref{eq:2feqs}, we see that, for any ordering of the one-forms $\omega_{k}$, which without loss of generality we take to be $1,2,\dots,n$, and label the other one-forms $\omega_{\alpha}$, $\alpha=n+1,\dots,m$, we have
\[
\begin{matrix*}[l]
  s'_{1}&\text{is the number of independent invariants}&I_{I;J1};\\
  s'_{2}&\text{is the number of independent invariants}&I_{I;J2};\\
  \hdotsfor{3}\\
  s'_{n}&\text{is the number of independent invariants}&I_{I;Jn};\\
  s'_{k}&\text{for $n<k\le m$ is equal to}&0.
\end{matrix*}
\]
In the above, when we say  for example ``the number of independent forms $I_{I;Jk}$'', we mean that the $I_{I;Jk}$ are not only independent among themselves, but also with respect to all $I_{I;Jl}$ for $l<k$. Thus, for any system of involutive seeds satisfying in addition condition \textbf{O1}, in particular, for any system of involutive seeds together with an involutive ordering, $s'_{i}$ is just the number of involutive seeds with the last index equal to $i$. 

However, the second version of involutive test concerns also the pseudo-character. We now show that the pseudo-character either vanishes or is equal to the non-vanishing $s'_{m}$. Observe that, for every independent invariant that is \emph{not} a seed there is an independent one-form equation. Hence $s'_{0}$ is the number of independent non-seed invariants minus $m$, as $m$ of these have been used implicitly to solve for the $\omega_{\mu}$. The above counting, on the other hand, shows that the number of seeds is given by $\sum_{i=1,m}s'_{i}$, so
\[
s'_{0}+s'_{1}+\dots+s'_{m}=M-m,
\]
where $M$ is the dimension of the space formed with all the independent invariants included in our system, showing that the pseudo-character and the last character are equal.

 Using the second version of the involutivity test, if we can show
 \begin{equation}
   \label{eq:invtes}
   s'_{1}+2s'_{2}+\dots+ns'_{n}=N
 \end{equation}
where $N$ is the number of free parameters for an integral element, then the system is involutive. To calculate the number of free parameters, we expand $dI_{I;Jk}$, which hitherto has been considered to be independent, in terms of $\omega_{k}$, $\omega_{\alpha}$. But this expansion is just the definition of the invariants $I_{I;Jkl}$:
\[
dI_{I;Jk}=I_{I;Jkl}\omega_{l}+\cdots
\]
where the omitted terms contain no free parameters, by our assumption. Hence $N$ is the total number of algebraically independent first derivations of the involutive seeds. Since we have included enough invariants in our original set, we see that the $I_{I;Jkl}$ are only subject to the derived relations and the generic relations. Hence an independent and spanning set of $I_{I;Jkl}$ can be taken as those $I_{I;Jkl}$ for which $k\ge l$ (using the generic relations) and $I_{I;Jk}$ is an involutive seed (using the derived relations).

Now, for a system of involutive seeds together with an involutive ordering, this, together with the condition \textbf{O2}, means that
\[
\begin{matrix*}[r]
  \text{The number of indep.}&I_{I;Jk1}&\text{is the number of seeds}&I_{I;J1},I_{I;J2},\dots,I_{I;Jn};\\
  \text{The number of indep.}&I_{I;Jk2}&\text{is the number of seeds}&I_{I;J2},\dots,I_{I;J,n};\\
  \hdotsfor{4}\\
  \text{The number of indep.}&I_{I;Jkn}&\text{is the number of seeds}&I_{I;J,n}.
\end{matrix*}
\]
Thus, \eqref{eq:invtes} is satisfied, and the system is involutive. Our interest lies in the degree of arbitrariness, i.e., the last non-zero Cartan character, and this is just the number of seeds with largest the last index. If this index is $k$, then the degree of arbitrariness $s_{k}=s'_{k}$ has the interpretation that the general solution of the system of invariants depends on $s_{k}$ free functions of $k$ variables.

A last very important thing that we need to note is that, by applying the theorem on the complete prolongation of Pfaffian systems, the system of invariants, which we have found to be involutive, can be prolonged at will, with the degree of arbitrariness unchanged: in particular, our system can be taken to include any given invariants.

Note that since the coordinates we use are provided by the invariants themselves, we do not have to worry about any degree of arbitrariness coming from the ``freedom'' in choosing the coordinates. This is what is meant by saying that our system is ``coordinate-free'': the price to pay being the condition \textbf{R}.

The claims of the ``directions of evolution'' follows directly from the interpretation of the Cartan--K\"ahler theorem.

\subsection{The original structural equations}

We have thus obtained a solution to the differential system for the invariants. With this solution, the one-forms $\omega_{k}$ and $\omega_{\alpha}$ can be expressed in terms of any set of invariants which we take to be the coordinates. It remains to show that the one-forms $\omega_{k}$ and $\omega_{\alpha}$ thus constructed satisfy the original structural equations of the problem, which we have ignored so far. Since the differential system we use to obtain the solution respects the independence condition, we can find a set of functionally independent invariants $I_{\mu}$, whose number is equal to the number of independent forms. For $I_{\mu}$, our differential system include the equation
\begin{equation}
\label{eq:reversemagick}  
dI_{\mu}=I_{\mu,\nu}\omega_{\nu}
\end{equation}
where our notation has changed so that $\omega_{\mu}$ includes both $\omega_{k}$ and $\omega_{\alpha}$. In this notation, our original structural equations are written
\[
d\omega_{\mu}=c_{\mu\nu\lambda}\omega_{\nu}\wedge\omega_{\lambda},
\]
which are \emph{not} included in our differential system, 
and the derivative is written as
\[
dc_{\mu\nu\lambda}=c_{\mu\nu\lambda,\gamma}\omega_{\gamma},
\]
which can all be included in our differential system due to the possibility of unlimited prolongation,
and the invariants $I_{\mu}$ are chosen among the $c_{\mu\nu\lambda}$ and its derivatives.

 As both $dI_{\mu}$ and $\omega_{\nu}$ are co-frames for the solution integral variety, the matrix $I_{\mu,\nu}$ is non-singular. Deriving \eqref{eq:reversemagick}, we get
 \begin{equation}
   \label{eq:aaba}
   0=dI_{\mu,\nu}\wedge\omega_{\nu}+I_{\mu,\nu}d\omega_{\nu}=I_{\mu,\nu\lambda}\omega_{\nu}\wedge\omega_{\lambda}
 \end{equation}
But as now we can take any invariant in the system we have used to solve for the degree of arbitrariness, $I_{\mu,\nu\lambda}$ is related to $I_{\mu,\nu}$: for example, as
\[
dc_{\mu\nu\lambda}=c_{\mu\nu\lambda,\gamma}\omega_{\gamma},
\]
we have
\[
0=d^{2}c_{\mu\nu\lambda}=c_{\mu\nu\lambda,\gamma\delta}\omega_{\gamma}\wedge\omega_{\delta}+c_{\mu\nu\lambda,\gamma}c_{\gamma\rho\delta}\omega_{\rho}\wedge\omega_{\delta},
\]
where $\omega_{\gamma\rho\delta}$ are the fundamental invariants. It is essential to note that this equation is a consequence of our differential system for the invariants and we do not need to assume the original structural equations: indeed, the algebraic relations for $c_{\mu\nu\lambda,\gamma\delta}$ in our differential system are required to include those relations using which this relation holds. Thus, \eqref{eq:aaba} can be written
\begin{equation}
  \label{eq:keyed}
  I_{\mu,\nu}(d\omega_{\nu}-c_{\nu\rho\lambda}\omega_{\rho}\wedge\omega_{\lambda})=0.
\end{equation}
As $I_{\mu,\nu}$ is non-singular, 
\[
d{\omega_{\nu}}-c_{\nu\rho\lambda}\omega_{\rho}\wedge\omega_{\lambda}=0,
\]
and the original structural equations are satisfied by our solution. Thus the proof of our algorithm is complete for the case where condition \textbf{R} holds and there are no additional functions.

\subsection{The case with non-maximal number of invariants}
\label{sec:nonmaxinv}

The problem with the case where the condition \textbf{R} fails is that, first, it is impossible to solve all of $\omega_{\mu}$ in terms of the differentials of the invariants, and thus it is not possible to form a differential system for the invariants in the way that we did for the non-singular case, and second, even if the first difficulty is somehow overcome, in \eqref{eq:keyed} the matrix $I_{\mu,\nu}$ is singular, and thus there is no way to ensure that the original structural equations are really satisfied.

Since in this case, among the differential of the invariants, there are only $\rho$, $0\le \rho <m$ linearly independent ones where $m$ is the total dimension, there are also only $\rho$ functionally independent invariants. To make progress, we first consider the equivalence problem for such a system. The method of equivalence is briefly summarised in appendix \ref{sec:meqv}.

For an equivalence problem formulated on the manifold $M$ of our system, let us consider the Euclidean space $\mathcal{C}$ of the invariants $c_{ijk}$ up to sufficiently high order so that all functionally independent invariants are guaranteed to be included. The exact order is immaterial, that we have considered a lot of redundant variables does not matter either. Then, for every concrete system, we have a {classifying map} $T:M\rightarrow \mathcal{C}$. This map defines a {classifying manifold} in the space of invariants. Using the classifying map and manifold, we can state the equivalence condition as follows: if two points $P$ and $\bar P$ of two systems defined on $M$ and $\bar M$ map to the same point in the classifying manifold (the classifying manifolds of the two problems are identified in the obvious manner), then the two systems are equivalent at the points $P$ and $\bar P$; if this condition holds for all points in the open sets $S\subset M$ and $\bar S\subset \bar M$, then the identification of $S$ and $\bar S$ that makes the condition holds provides an equivalence of the two systems.

For our present purpose, we are interested in the case where $M=\bar M$. Let us focus on the image of the classifying map $T:M\rightarrow \mathcal{C}$. At regular points, this image is a submanifold of $\mathcal{C}$: its dimension is exactly $\rho$. If $\rho < m$, then the pre-image of a point $Q\in \mathcal{C}$ is non-trivial: $T^{-1}(Q)$ is locally a $m-\rho$ dimensional submanifold in $M$. But if we set up the equivalence problem in trying to deduce the equivalence of $M$ with itself, but identifying a point $P$ in $M$ with a nearby point $P'$ in a neighbourhood, we see from the reasoning of the previous discussion that as long as $P$ and $P'$ are in the same pre-image, i.e., as long as $T(P)=T(P')$, the equivalence problem has a solution. As $P$ and $P'$ can be connected by a path not going out of the pre-image, we see that this self-equivalence is actually a \emph{symmetry} under a finte dimensional Lie group, and $m-\rho$ gives the dimension of the symmetry group of the problem.
 (If $\rho=m$, then no non-trivial continuous symmetry group can exist.)
Thus, for systems with a non-maximal number $\rho$ of invariants, we can find $m-\rho$ vector fields $\mathbf{v}_{\alpha}$, corresponding to the symmetry group, for which
\[
\mathcal{L}_{\mathbf{v}}\omega_{\mu}=0,\qquad \mathbf{v}(c_{ijk,lm\dots})=0.
\]

Now suppose we take a submanifold $N$ of dimension $\rho$ transverse to all of the the vectors $\mathbf{v}_{\alpha}$, and we find a solution of our system (i.e., a functional dependence of the forms $\omega_{\mu}$ and the invariants $c_{\mu\nu\rho}$ on $m$ coordinates $x_{\mu}$ satisfying the structural equations) valid in an infinitesimal neighbourhood of $N$, then using the system of vector fields $\mathbf{v}_{\alpha}$, this solution can be extended to the whole space.
Also, any solution valid for the whole space $M$, when restricted to such a transverse submanifold $N$, will also satisfy the structural equations: the equations that are to be satisfied now are just the pullbacks of the equations on the total space, and exterior differentiation commutes with pullbacks. On the other hand, the converse is in general not true: for simplicity, let $x_{1},x_{2},\dots,x_{\rho}$ be the coordinates on the submanifold $N$, and $x_{\rho+1},\dots,x_{m}$ be the transverse coordinates. Then our structural equations are written in the differentials of these coordinates. That we have a solution on $N$ means that, for the structural equation,
\[
d\omega_{\mu}(\pd_{x_{\alpha}},\pd_{x_{\beta}})=c_{\mu\nu\rho}\omega_{\nu}\wedge\omega_{\rho}(\pd_{x_{\alpha}},\pd_{x_{\beta}}),\qquad(\alpha,\beta\le \rho)
\]
is satisfied on at points on $N$. We can even show that for certain coordinates and for $\alpha,\beta>\rho$, this equation also holds, due to the action of the Lie group. However, in general, when $\alpha\le\rho$ and $\beta>\rho$, there is no reason that this equation will hold. Hence, that the structural equations are satisfied when pulled back onto any submanifold $N$ thus chosen is a necessary, but in general not sufficient condition for a solution of the original equations.

Now let us return to our differential invariants and involutive seeds. First observe that, if $I_{I;Jk}$ is an involutive seed, then $\omega_{k}$ depends on the differentials of the invariants: indeed,
\[
dI_{I;J}=I_{I;Jk}\omega_{\bar k}+\cdots
\]
and due to the requirements of the involutive seeds, when we prolong the problem, 
\[
dI_{I;Jk}=I_{I;Jk\bar k}\omega_{\tilde k}+\cdots
\]
where $I_{I;Jk\bar k}$ is an \emph{independent} invariant. This shows that for all solutions, we can find a submanifold $N$ such that all the forms $\omega_{k}$ occurring with a differential of the involutive seed in equation \eqref{eq:2feqs} remain independent. On other other hand, clearly if for the system \eqref{eq:2feqs}, if we take any $m-\rho$ forms $\omega_{\mu}$ other than those $\omega_{k}$ occurring explicitly with the involutive seeds in \eqref{eq:2feqs} to be forms written in terms of the invariants, with \emph{arbitrary} functional dependence, we can now solve the remaining $\rho$ one-forms $\omega_{\mu}$, which contains the $\omega_{k}$, in terms of the invariants. Due to the form of the equation \eqref{eq:2feqs}, for any such choice, the Cartan characters for this system, with $\rho$ independence conditions, are the same, and by a reasoning exactly the same as in the maximal rank case we see that this system is involutive, with degree of arbitrariness given by the last non-zero Cartan character. Observe also that if for a system of involutive seeds and ordering, if the maximal last index is $k$, then $\rho\ge k$: this can be seen easily from \eqref{eq:2feqs} as well. This means that the degree of arbitrariness always occurs at a dimension $\le \rho$.

Using the Lie group action, such a system of values of the invariants can be extended to the whole space. But as remarked earlier, there is no guarantee that after extension all of the structural equations will be satisfied, hence in this case we have obtained only an upper bound.

Finally, let us remark that for the following special case, the upper bound is realised: the structural equations reads
\[
\left\{
  \begin{aligned}
    d\omega_{\alpha}&=C_{\alpha\beta\gamma}\omega_{\beta}\wedge\omega_{\gamma},\\
    d\omega_{i}&=c_{ijk}\omega_{j}\wedge\omega_{k},
  \end{aligned}
\right.
\]
where $C_{\alpha\beta\gamma}$ are constants, and all of $\omega_{i}$ can be solved in terms of $c_{ijk}$ and their invariants: it suffices to first ignore the first set of equations and obtain the degree of arbitrariness for this system. This degree of arbitrariness is realised since the first set of equations is consistent (otherwise the Bianchi relations will have something that reduces to $1=0$), and from the theory of Lie groups we know that there exists a Lie group satisfying the first set of equations. The solution space is then the product space formed by the Lie group and the solution whose degree of arbitrariness we know. As an easy consequence, for the following system the upper bound is also realised:
\[
\left\{
  \begin{aligned}
    d\omega_{\alpha}&=C_{\alpha\beta\gamma}\omega_{\beta}\wedge\omega_{\gamma}\pmod{d\omega_{i}},\\
    d\omega_{i}&=c_{ijk}\omega_{j}\wedge\omega_{k},
  \end{aligned}
\right.
\]
where except for the modulus part, the assumptions are the same as before.

\subsection{Additional functions}

We only consider the case satisfying condition \textbf{R}, as our algorithm states. In this case, we simply adjoin the additional functions and their derivatives to the system \eqref{eq:2feqs}, and the result is obvious. The condition \textbf{I3} is necessary since only those functions satisfying \textbf{I3} will appear in the written out part in the right hand side of \eqref{eq:2feqs}, for which the Cartan characters are calculated.

\appendix
\section{Cartan--K\"ahler theory, involutivity and prolongation}
\label{sec:ckip}

\paragraph{Exterior differential systems.}

Let a exterior differential system be defined on a manifold $M$, formed by
\begin{align*}
  &\text{functions:}&&f_{1},\dots,f_{r_{0}},\\
  &\text{one-forms:}&&\theta_{1},\dots,\theta_{r_{1}},\\
  &\text{two-forms:}&&\Theta^{(2)}_{1},\dots,\Theta^{(2)}_{r_{2}},\\
  &\text{three-forms:}&&\Theta^{(3)}_{1},\dots,\Theta^{(3)}_{r_{2}},\\
  &\cdots&&\cdots
\end{align*}
We also assume that that the given system has already been transformed into a closed system:
\[
d\theta_{i}=0\pmod{\Theta_{1}^{(2)},\dots,\Theta_{r_{2}}^{(2)}},\qquad\text{etc.}
\]
so all relations that can be obtained by differentiation have already been incorporated. A solution (integral variety) of this exterior differential system is a submanifold of $M$ such that all the forms in the system vanish when restricted to this submanifold.

\paragraph{Characteristics.}

Before talking about Cartan--K\"ahler theorem, we need to talk about {characteristic directions}. A direction given by the vector $\mathbf{v}$ is characteristic if
\[
\mathbf{v}\iprod \Theta=0\qquad\text{for all }\Theta\text{ in the differential system},
\]
i.e., characteristics directions are the ones having the property that if an element contains this direction, then it is automatically an integral element. To obtain the integral varieties {of the characteristics}, define the {characteristic system} $\mathcal{C}$ associated with our original differential ideal, consisting of all one-forms $\vartheta$ such that
\[
\vartheta(\mathbf{v})=0\qquad\text{for all }\mathbf{v}\in\mathcal{V}.
\]
Of course, not all systems contain characteristic directions.

The characteristic system is important when we are solving the Cauchy problem. Suppose we are integrating a differential system and we want to go from dimension $k$ to dimension $k+1$. We need to specify the initial data on the integral variety of dimension $k$ which we have already found. But if this variety contains a characteristic variety, then we know two things: first, the data along these varieties are well-defined when we know them {at any point} on them, hence there are consistency issues when specifying the initial data; second, we need specify extra functions in order to effect the integration. Hence for the Cauchy problem, specifying initial data on varieties containing characteristic variaties is problematic and in general inconsistent.

In the following, we assume that any vectors and any submanifold that we talk about do not contain characteristic directions.

\paragraph{Integral elements; Cartan characters.}

A first step in constructing integral variety is constructing integral elements: these are subspaces of the tangent space at a single point which satisfies the equations. We construct the integral elements recursively. Obviously, it suffices to restrict our attention to integral elements over the submanifold defined by the vanishing of the functions in our differential system. 

First let us consider {integral elements at a single point}. Every one-dimensional integral element $\mathbf{v}$ must satisfy
\[
\theta_{i}(\mathbf{v})=0.
\]
Hence the space of integral element of dimension $1$ is given by a linear equation: $A_{ij}$ are constants at the points we consider and any solution $v_{j}$ forms just a vector. Now suppose we already have a determined $1$ dimensional linear element $\mathbf{v}_{1}$ and we would like to extend it into a $2$ dimensional linear element: this amounts to finding another direction $\mathbf{v}_{2}=\bar v_{i}\mathbf{I}_{i}$. We need to ensure $A_{ij}\bar v_{j}=0$: this direction must itself be a solution to the one-dimensional problem. But there is more: suppose the $2$-forms in the system are
\[
\Theta_{i}^{(2)}=A_{ijk}\pi_{j}\wedge\pi_{k},\qquad A_{ijk}=-A_{ikj},
\] 
then we must have
\[
A_{ijk}v_{j}\bar v_{k}=0.
\]
$v_{j}$ is data already given to us: this, together with $A_{ij}\bar v_{j}=0$, forms a linear system whose solution space gives all possible directions extending $\mathbf{v}_{1}$. In general, given a set $\mathbf{v}_{1}$, $\dots$, $\mathbf{v}_{p}$ forming an integral element of dimension $p$, the extension to dimension $p+1$ is obtained by a suitable linear system on the free tangential directions. This linear system is deduced from all the exterior forms in the system of dimension up to $p+1$. As we go up in the dimension of the integral element, the dimension of the solution space will decrease: for every step it will decrease by at least one: for example, at the second step, $\mathbf{v}_{1}$ and $\mathbf{v}_{2}$ must be independent for them to constitute an extension of $\mathbf{v}_{1}$. Thus we will eventually come to a dimension where the integral element can no longer be extended.

The integral varieties we are looking for have something to do with the ranks of the various linear systems we have just described. But we cannot ensure that, for all choices of the point $P$ in $M$, the rank of the system $A_{ij}v_{j}$ remains the same.
Thus let us call a point $P$ an {integral point} if it is on the algebraic variety defined by the vanishing of functions $f_{i}=0$ in the differential system. An integral point $P$ is {generic}
 if, at the point, $df_{i}=f_{i,j}dx^{j}$ has maximal rank in a neighbourhood. This ensures that we are not dealing with some pathological algebraic variety.

On a generic point, let us find the one-dimensional integral elements $\mathbf{v}$.
If the rank of $A_{ij}$ defining the integral element is maximal {in a neighbourhood}, then the integral element is said to be {ordinary} and {this generic point} (not the integral element!) is said to be a {regular} point. The rank of the matrix $A_{ij}$ is called the {zeroth Cartan character} (or simply zeroth character) of the system at the point $P$ and is denoted $s_{0}$.

Now given an ordinary one-dimensional integral element $\mathbf{v}_{1}$ at a point, let us try to extend it by one dimension. This element is said to be {regular} if the rank of the system $A_{ij}\bar v_{j}$, $A_{ijk}v_{j}\bar v_{k}$, where $v_{j}$ is now given, is maximal in a neighbourhood. The solutions for such a maximal-rank system $\mathbf{v}_{1}$, $\mathbf{v}_{2}$, are said to be {ordinary} integral elements. Since the rank of the system $A_{ij}\bar v_{j}$, $A_{ijk}v_{j}\bar v_{k}$ cannot be less than the rank of a part of itself, $A_{ij}\bar v_{j}$, we denote the rank as $s_{0}+s_{1}$. The integer $s_{1}$ is called the {first Cartan character}.

 The general pattern should be clear: for the linear system defined at a particular dimension $p$, the maximality of its rank defines the {ordinary} elements at this dimension, and the {regular} elements at one less dimension. Then, as we have remarked that the dimension of the integral elements cannot increase indefinitely, we will come to a dimension $p$ where we have ordinary elements that cannot be further extended, hence there are no regular elements at this dimension. The rank of this final system is
\[
s_{0}+s_{1}+\dots+s_{p},
\]
and for such a system we have a set of $p+1$ Cartan characters.

After this long train of definitions, we can finally state

\begin{ckt}
  On a manifold of total dimension $n$, for a given integral variety
  admitting ordinary integral element of dimension $k$ at a point, we
  have
  \begin{equation}
    \label{eq:m.1}  
    s_{0}+s_{1}+\dots+s_{k}\le n-k,
  \end{equation}
  equality holds above if and only if $k$ is the largest such
  dimension. If all data used to define the exterior differential
  system are analytic functions, then the above condition is also
  sufficient: i.e.~there will exist a regular $k$-dimensional integral variety for the differential system.
\end{ckt}

Let us also define the Cartan {pseudo-character} $\sigma_{k}$ by the formula
\[
s_{0}+s_{1}+\dots+s_{k-1}+\sigma_{k}=n-k.
\]
It is always non-negative by the theorem, and if $k$ is the maximal dimension, $\sigma_{k}=s_{k}$.

\begin{rmk}
  The sufficiency part of the Cartan--K\"ahler theorem requires analyticity. For physical applications, should this worry us?

First, let us note that it is sometimes possible to substitute the Cauchy--Kowalewski theorem for some stronger existence theorem in the theory of partial differential equations: thus the Cartan--K\"ahler theorem actually holds in the smooth setting (as opposed to the analytic setting) under some circumstances. In particular this is true for involutive {hyperbolic} systems, quasilinear systems, and some Pfaffian systems. Most of the equations that arise in physics fall into one of these classes, so we actually have guaranteed $C^{\infty}$ results.

Second, as the necessity part does not require analyticity (neither does the ``degenerate'' case where the theorem reduces to Frobenius theorem), we will never miss any solutions by applying the Cartan--K\"ahler theorem in cases where the $C^{\infty}$ version does not hold: the worst we do is claiming the existence of solutions where there are none. But as the space of analytic functions is dense in the space of continuous functions, there is no physical ground for claiming that a function is differentiable to a high order but is not analytic: physically we cannot really distinguish, say,  $C^{4}$ data from $C^{\omega}$ data since the error in our knowledge would already make them indistinguishable. Furthermore, if we are really in a situation where such issues becomes significant, for example due to the discontinuity of the underlying space, maybe with origin in some kinds of ``quantum'' effects, there should be a lot more other things to worry about than analyticity.\hfill\P
\end{rmk}

\paragraph{Interpretation of the characters.}

For our purpose, the following interpretation of the theorem is important: the regular $k$-dimensional integral variety guaranteed by the theorem can be integrated by specifying
\begin{alignat*}{3}
  s_{1}+s_{2}+\dots+s_{k-1}+\sigma_{k}&\text{ arbitrary functions of }&&x^{1},\\
  s_{2}+\dots+s_{k-1}+\sigma_{k}&\text{ arbitrary functions of }&&x^{1},x^{2},\\
  \cdots&&&\cdots\\
  s_{k-1}+\sigma_{k}&\text{ arbitrary functions of }&&x^{1},x^{2},\dots,x^{k-1},\\  
  \sigma_{k}&\text{ arbitrary functions of }&&x^{1},x^{2},\dots,x^{k-1},x^{k}.
\end{alignat*}
where $x^{1}, x^{2},\dots$ is a coordinate system system for the manifold. Furthermore, the last non-zero (pseudo)-character is not affected by our choice of order of integration, the form we used to write down the differential system, etc.~(the characters before that are affected by such choices). Obviously, for a differential system, the last non-zero (pseudo)-character is its degree of arbitrariness when we are interested in forming $k$-dimensional integral varieties, and the index of this (pseudo)-character is the dimension at which this degree of arbitrariness occurs.
 
\paragraph{Independence conditions; involutive systems.}

The solutions guaranteed by the Cartan--K\"ahler theorem has a problem: it does not care for our requirement of {independent variables}. To make progress, let us define a differential system with independence condition as a differential system for which we require that solutions must keep certain one-forms $\omega_{1}$, $\omega_{2}$, $\dots$, $\omega_{m}$ independent. This amounts to 
\[
\omega_{i}\wedge\omega_{j}\wedge\dots\wedge\omega_{k}\neq 0,\qquad i,j,\dots,k\text{ all distinct}
\]
on solutions for any choice of any numbers of $\omega_{i}$, $\omega_{j}$, $\dots$. For the coordinates $x_{1}$, $x_{2}$, $\dots$, $x_{m}$, we just take the forms to be $dx_{1}$, $dx_{2}$, $\dots$, $dx_{m}$, and the vectors spanning the integral elements we look for must be of the form
\begin{equation}
  \label{eq:ordintelem}
  \frac{\pd}{\pd x_{a}}+\sum_{i=m+1}^{n} B^{i}_{a}\frac{\pd}{\pd z_{i}},\qquad{a=1,\dots,m,}
\end{equation}

Immediately, we see that we \emph{must not} end up with equations of the form
\begin{equation}
  \label{eq:esstor}
  0=\Omega^{(k)}=A_{ij\dots k}\,\omega_{i}\wedge\omega_{j}\wedge\dots\wedge\omega_{k}.
\end{equation}
Unless $A_{ij\dots k}=0$, no solution will satisfy the independence conditions. Such terms are called {essential torsion}. To proceed in such cases, we need to add to our differential system the algebraic equations
\[
A_{ij\dots k}=0.
\]
Such equations might have no solution: for example if $A_{ij\dots k}$ are non-zero constants. We say that such a differential system is {incompatible}.

Now consider only systems with no essential torsion.
Our aim is to obtain solutions satisfying the independence conditions as {general solutions} of the problem without independence conditions. Hence, we will call differential systems for which the constraints on ordinary $m$ dimensional integral elements {do not require any linear relations} among the independence conditions $\omega_{1}$, $\dots$, $\omega_{m}$ \emph{involutive systems}. This does not mean that all ordinary $m$ dimensional integral elements satisfy the independence conditions: it means that \emph{almost all} satisfy, and \emph{almost all} integral elements spanned by the vectors of the form \eqref{eq:ordintelem} are ordinary, since the conditions for otherwise are both equality conditions. For involutive systems, we can obtain the integral varieties we want by applying the Cartan--K\"ahler theorem to ordinary integral elements of the form \eqref{eq:ordintelem}.

\paragraph{Reduced characters; Cartan's tests.}

The requirement that none of the constraints on integral elements can involve any relations for $\omega_{i}$ means that, in calculating the Cartan characters, we can ignore all directions that correspond to $\omega_{i}$ {For example, we have two forms \[\Omega^{\alpha}=A^{\alpha}_{ij}\omega_{i}\wedge\omega_{j}+B^{\alpha}_{ia}\omega_{i}\wedge\pi_{a}+C^{\alpha}_{ab}\pi_{a}\wedge\pi_{b}\] and in calculating the Cartan character, the first direction is chosen as the vector $\mathbf{I}_{1}$ dual to $\omega_{1}$. With this direction, we have
\[
\Omega^{\alpha}(\mathbf{I}_{1})=A^{\alpha}_{1j}\omega_{j}+B^{\alpha}_{1a}\pi_{a}.
\]
The requirement that we must not have any constraints for the independent directions (i.e., no relations of the form $c_{i}\omega_{i}=0$) means that the rank of the system $(A^{\alpha}_{1j},B^{\alpha}_{ia})$ is the same as the rank of the system $(B^{\alpha}_{ia})$. Notice we do this only {after} using the already found tangent direction.

Let us call the numbers
\[
s'_{0},\qquad s'_{1},\qquad s'_{2},\qquad \dots\qquad s'_{m-1}
\]
which are calculated by omitting all terms corresponding to $\omega_{i}$ in the calculation for Cartan characters the {reduced Cartan characters}:\index{reduced Cartan character} this definition holds for both involutive and non-involutive system. The {reduced pseudo-character} $\sigma'_{m}$ is defined as
\[
s'_{0}+s'_{1}+\dots+s'_{m-1}+\sigma'_{m}=n-m.
\]

Involutive system has the normal characters and reduced characters equal. Conversely, if all reduced characters are equal to the non-reduced counterparts, this implies there is no relation among the independent variables, and hence the system is involutive. We therefore have a necessary and sufficient condition for involutive systems: the equality of the reduced and normal characters.

It is troublesome to calculate two sets of characters, especially the non-reduced ones. The following theorem gives a simple way to tell if a differential system with independence condition is involutive:
\begin{cit}
  A differential system with independence condition is involutive if and only if
  \begin{equation}
  \label{eq:invthm}
  N=s'_{1}+2s'_{2}+\dots+(k-1)s'_{k-1}+k\sigma'_{k},
\end{equation}
where $N$ is the total number of free parameters occurring for the ordinary $k$-dimensional integral element.
\end{cit}
Calculating reduced characters is easier than calculating the normal characters, but we still need to ensure that the integral element we specify is generic, meaning that the rank of the various systems are maximal. Now assume that we calculate the reduced characters by ignoring the step of checking the rank condition. Then we still have
\begin{cit2}
  A differential system with independence condition is involutive if 
  \begin{equation}
  \label{eq:invthm}
  N=s'_{1}+2s'_{2}+\dots+(k-1)s'_{k-1}+k\sigma'_{k},
\end{equation}
where there various reduced characters are calculated without checking the rank condition, and $N$ is the total number of free parameters occurring for the ordinary $k$-dimensional integral element.
\end{cit2}
Note that there is no ``only if'' for this easier test. In practice, for calculating the number of free parameters, we take the differential one-forms $\varpi_{\alpha}$ that are not required to be independent in our system, and write
\[
\varpi_{\alpha}=A_{\alpha i}\omega_{i}
\]
where $\omega_{i}$ are the independent forms. Then substitution into the system gives algebraic constraints on $A_{\alpha i}$. The components that are not constrained gives the number of free parameters.

\paragraph{Prolongation.}

Actually, when formulating the original problem, we can also take the some of the parameters $A_{\alpha i}$ occurring above as variables. Doing this is called effecting a prolongation of a differential system (by addition of new variables). If we take all of the parameters as new variables, we are effecting a complete prolongation. The Cartan--Kuranishi theorem states that for well-behaving systems (which we will not make precise here), after a sufficient number of complete prolongations, we either arrive at an involutive system or an incompatible system. For us, on the other hand, we need the following theorem: 
\begin{cpt}
  If a differential system consists only of functions, one-forms, and two-forms that are generated by differentiating the one-forms (i.e., a Pfaffian system), and the system is involutive, then its complete prolongation is also involutive, with the Cartan characters of the new system related to the old system by
\begin{alignat*}{5}
  s_{1}\str&=&s_{1}+s_{2}+\dots+s_{m-1}+s_{m},\\
  s_{2}\str&=&s_{2}+\dots+s_{m-1}+s_{m},\\
  &\dots&\\
  s_{m-1}\str&=&s_{m-1}+s_{m},\\
  s_{m}\str&=&s_{m}.
\end{alignat*} 
\end{cpt}

This theorem also explicitly verifies the general result that the last non-zero Cartan character is an intrinsic property of the system, i.e., not affected by the way of presentation.

\section{The method of equivalence}
\label{sec:meqv}

Our task is to understand, given two co-frames $\theta_{1}$, $\dots$, $\theta_{n}$ and $\bar\theta_{1}$, $\dots$, $\bar\theta_{n}$ defined on two differential manifolds $M$ and $\bar M$ of dimension $n$, when they are ``locally the same'', meaning there is a map $f:M\rightarrow\bar M$ such that
\[
f^{*}\bar\theta_{i}=\theta_{i}.
\]
We will omit pullback signs from now on and just write $\theta_{i}=\bar\theta_{i}$.

We can treat this equivalence problem as the integrability of the form
\[
\vartheta_{i}=\bar\theta_{i}-\theta_{i}
\]
in the space $\bar M\times M$. We are interested in the $m$-dimensional integral manifolds which are {transverse}, meaning that either $\theta_{i}$ or $\bar\theta_{i}$ can be taken as the independent conditions.

As we are dealing with coframes, we can calculate their exterior derivatives
\[  d\theta_{i}=c_{ijk}\theta_{j}\wedge\theta_{k},\qquad
  d\bar\theta_{i}=\bar c_{ijk}\bar\theta_{j}\wedge\bar\theta_{k},
\]
so
\[d\vartheta_{i}
=\bar c_{ijk}\bar\theta_{j}\wedge\vartheta_{k}+\bar c_{ijk}\vartheta_{j}\wedge\theta_{k}+(\bar c_{ijk}-c_{ijk})\theta_{j}\wedge\theta_{k}
\]
The existence of integral manifolds then requires
\begin{equation}
  \label{eq:m.3}
  \bar c_{ijk}=c_{ijk},
\end{equation}
i.e., the system must be Frobenius integrable (integral manifolds from the the Cartan--K\"ahler theorem will never be of the required dimension, since we already have $s_{0}=2m-m=m$, and any non-zero $s_{i}$ for $i\neq 0$ will lower the dimension of the integral manifold).

The simplest case is where $\bar c_{ijk}$ and $c_{ijk}$ are both constants. If they are equal, then the two systems are equivalent, otherwise (including the case where one set are constants while the other are not) they are not.

If $\bar c_{ijk}$ and $c_{ijk}$ are not constants, then we require $f^{*}\bar c_{ijk}=c_{ijk}$, but for functions under diffeomorphism we cannot directly compare them. Worse, even though we can implicitly define the submanifold of $\bar M\times M$ by the relation $\bar c_{ijk}=c_{ijk}$, it is far from certain that restricted to this submanifold what will happen to $\theta_{i}$ and $\bar\theta_{i}$, which we require to be independent. 

To make progress, we treat \eqref{eq:m.3} as a new condition and adjoin it to our conditions for equivalence (prolongation). The closure of our system now includes the condition
\[
d\bar c_{ijk}=d c_{ijk},\qquad
\text{or}\qquad
\bar c_{ijk;l}=c_{ijk;l},
\]
where the {coframe derivative} for a function $h$ is defined, as usual,
\[
dh= h_{;i}\theta_{i}.
\]

Let us first investigate the case where $c_{ijk}$ and $\bar c_{ijk}$ both contain $m$ independent functions among them each. Let us denote these by
\[
I_{1},\quad I_{2},\quad\dots\quad I_{m};\qquad \bar I_{1},\quad \bar I_{2},\quad\dots\quad \bar I_{m},
\]
and the rest of the quantities are expressible as functions of them. Being functional independent, their differentials are linearly independent, or the matrix $I_{i;j}$ in the following expression is invertible:
\[
dI_{i}=I_{i;j}\theta^{j}.
\]
In this case, $c_{ijk}=\bar c_{ijk}$ implies $I_{i}=\bar I_{i}$, which in turn implies $d I_{i}=d\bar I_{i}$ and hence $\theta_{i}=\bar\theta_{i}$. Observe that we do not even need to check the equality for the terms other than $I_{i}$. Indeed, let us differentiate the above equation. We get:
\[
0=I_{i;jk}\theta^{k}\wedge\theta^{j}+I_{i;j}c_{jkl}\theta^{k}\wedge\theta^{l}
\]
so $I_{i}=\bar I_{i}$ implies $\theta_{i}=\bar\theta_{i}$ and in turn these two equations together implies the equality of everything else. Hence, in this case, we only need to check
\[
I_{i}=\bar I_{i},\qquad I_{i;j}=\bar I_{i;j}.
\]
In the general case, what we do is we use the coframe derivative to differentiate the \emph{fundamental invariants} $C_{ijk}$ until we get no more functional independent quantities. For example, in this way we may obtain a system
\[
  \bar\theta_{i}=\theta_{i},\qquad
  \bar c_{ijk}=c_{ijk},\qquad
  \bar c_{ijk;l}=c_{ijk;l},\qquad
  \bar c_{ijk;lm}=c_{ijk;lm},\qquad
  \bar c_{ijk;lmn}=c_{ijk;lmn},
\]
where $c_{ijk;lmn}$ introduce no new functionally independent quantities. These conditions are obviously necessary. They are also sufficient for equivalent, as it can be easily checked that they imply the Frobenius integrability of the system
\[
  \bar\theta_{i}=\theta_{i},\qquad
  \bar c_{ijk}=c_{ijk},\qquad
  \bar c_{ijk;l}=c_{ijk;l},\qquad
  \bar c_{ijk;lm}=c_{ijk;lm},
\]
and the integral manifold really is transverse.

\newpage
\nocite{Alvarez:2009p4142}
\nocite{JOlver:1995p9015,LBryant:1991p8950}
\nocite{cartan-oc}
\nocite{cartan-eds}
\nocite{cartan-newtonian}
\nocite{cartan-einstein}
\nocite{hu1}
\nocite{hu2}
\nocite{edsym}
\nocite{mdmb}
\nocite{mdm-o}
\nocite{Giulini:2006p66}
\nocite{pirani1962}
\nocite{Alvarez:2007p4209}
\nocite{meq}
\nocite{Estabrook:1964p2608}
\nocite{BONeill:1966p7043}
\nocite{Dyadic Analysis of Space-Time Congruences}
\nocite{Wahlquist:1967p2556}
\nocite{cartanequ}
\nocite{tensor}
\nocite{WSharpe:1997p5521}
\nocite{Wahlquist:1966p2548}
\nocite{cartan-pf}
\nocite{cartan-grp}

\bibliographystyle{hplain}

\begin{thebibliography}{10}

\bibitem{Alvarez:2007p4209}
Orlando Alvarez.
\newblock Schwarzschild spacetime without coordinates.
\newblock {\em arXiv}, Jan 2007, gr-qc/0701115v2.

\bibitem{Alvarez:2009p4142}
Orlando Alvarez.
\newblock Black holes without coordinates.
\newblock {\em arXiv}, Apr 2009, 0904.0733v1.

\bibitem{born1909}
M~Born.
\newblock {\em Ann.~der Physik}, 30, 1909.

\bibitem{LBryant:1991p8950}
Robert~L. Bryant, S.S. Chern, Robert~B. Gardner, Hubert~L. Goldschmidt, and
  P.A. Griffiths.
\newblock {\em Exterior differential systems}.
\newblock Springer, Jan 1991.

\bibitem{cartan-pf}
\'Elie Cartan.
\newblock Sur l'int\'egration des syst\`emes d'\'equations aux
  diff\'erentielles totales.
\newblock {\em Ann.~Sci.~\'Ecole~Norm.~Sup.}, 18:241--311, 1901.

\bibitem{cartan-grp}
\'Elie Cartan.
\newblock Sur la structure des groupes infinis de transformations.
\newblock {\em Ann.~Sci.~\'Ecole~Norm.~Sup.}, 3(21), 1904.

\bibitem{cartan-phy}
\'Elie Cartan.
\newblock Sur la th\'eorie des syst\`emes en involution et ses applications \`a
  la relativit\'e.
\newblock {\em Bull.~Soc.~Math.~France}, 59:88--118, 1931.

\bibitem{cartanequ}
\'Elie Cartan.
\newblock Les probl\`emes d'\'equivalence.
\newblock {\em S\'eminaire de Math\'ematiques}, 1936.

\bibitem{cartan-eds}
\'Elie Cartan.
\newblock {\em Les systm\`emes diff\'erentiels ext\'erieurs et leurs
  applications g\'eom\'etriques}.
\newblock Hermann, 1945.

\bibitem{cartan-oc}
\'Elie Cartan.
\newblock {\em \OE{}uvres compl\`etes}.
\newblock Gauthier-Villars, 1953.

\bibitem{cartan-newtonian}
\'Elie Cartan.
\newblock Sur les vari\'et\'es \`a connexion affine et la th\'eorie de la
  relativit\'e g\'en\'eralis\'ee.
\newblock {\em Ann.~Sci.~\'Ecole~Norm.~Sup.}, 40(325):325--412, 1955.

\bibitem{cartan-n-t}
\'Elie Cartan.
\newblock {\em On manifolds with an affine connection and the theory of general
  relativity}.
\newblock Bibliopolis, 1986.

\bibitem{mdm-o}
Yvonne Choquet-Bruhat.
\newblock Th\'eor\`eme d'existence pour certains syst\`emes d'\'equations aux
  d\'eriv\'ees partielles non lin\'eaires.
\newblock {\em Acta Mathematica}, 1952.

\bibitem{mdmb}
Yvonne Choquet-Bruhat.
\newblock {\em General relativity and the Einstein equations}.
\newblock Oxford University Press, 2009.

\bibitem{cartan-einstein}
Robert Debever, editor.
\newblock {\em {\'Elie Cartan--Albert Einstein letters on absolute parallelism,
  1929-1932}}.
\newblock Princeton University Press, 1979.

\bibitem{Estabrook:1964p2608}
F~Estabrook and H~Wahlquist.
\newblock Dyadic analysis of space-time congruences.
\newblock {\em Journal of Mathematical Physics}, Jan 1964.

\bibitem{edsym}
FB~Estabrook.
\newblock Exterior differential systems for {Yang-Mills} theories.
\newblock {\em arXiv}, (0809.2174), 2008.

\bibitem{meq}
Robert~B. Gardner.
\newblock {\em The Method of Equivalence and Its Applications}.
\newblock Society for Industrial Mathematics, 1987.

\bibitem{Giulini:2006p66}
Domenico Giulini.
\newblock Algebraic and geometric structures of special relativity.
\newblock {\em arXiv}, Feb 2006, math-ph/0602018v2.

\bibitem{hu2}
Ziyang Hu.
\newblock The degree of arbitrariness of general very special relativity with
  holonomy constraint.
\newblock In preparation, 2011.

\bibitem{hu1}
Ziyang Hu.
\newblock A framework for structure-preserving submersions and its applications
  in physics.
\newblock In preparation, 2011.

\bibitem{AndrewIvey:2003p7876}
Thomas~Andrew Ivey and J~M. Landsberg.
\newblock {\em Cartan for beginners: differential geometry via moving frames
  and exterior differential systems}.
\newblock Jan 2003.

\bibitem{JOlver:1995p9015}
Peter~J. Olver.
\newblock {\em Equivalence, invariants, and symmetry}.
\newblock Cambridge University Press, Jan 1995.

\bibitem{BONeill:1966p7043}
B~O'Neill.
\newblock The fundamental equations of a submersion.
\newblock {\em Michigan Math. J}, Jan 1966.

\bibitem{pirani1962}
F~A~E Pirani and G~Williams.
\newblock Rigid motion in a gravitational field.
\newblock {\em S\'e{}minaire Janet}, 5, 1962.

\bibitem{WSharpe:1997p5521}
Richard~W. Sharpe.
\newblock {\em Differential geometry: Cartan's generalization of Klein's
  Erlangen program}.
\newblock Springer, Jan 1997.

\bibitem{tensor}
A~H Thompson.
\newblock The conformal generalisation of the {Herglotz--Noether} theorem.
\newblock {\em Tensor, N.S.}, 19, 1968.

\bibitem{Wahlquist:1966p2548}
H~Wahlquist and F~Estabrook.
\newblock Rigid motions in {E}instein spaces.
\newblock {\em Journal of Mathematical Physics}, Jan 1966.

\bibitem{Wahlquist:1967p2556}
H~Wahlquist and F~Estabrook.
\newblock Herglotz--{N}oether theorem in conformal space-time.
\newblock {\em Journal of Mathematical Physics}, Apr 1967.

\end{thebibliography}

\addcontentsline{toc}{section}{References}
\end{document}